\documentclass[11pt]{article}

\usepackage[margin=1.2in]{geometry}
\usepackage{microtype}
\usepackage{amsmath,amssymb,amsthm,mathtools,mathrsfs}
\usepackage{graphicx,booktabs,xcolor,enumitem,float}
\usepackage{tcolorbox}
\usepackage{algorithm}
\usepackage[noend]{algpseudocode}
\usepackage{tikz}
\usepackage{subcaption}
\usetikzlibrary{arrows.meta,calc,positioning,shapes.geometric}
\usepackage[normalem]{ulem}
\usepackage[style=authoryear,natbib=true,backend=biber,
            maxcitenames=2,mincitenames=1,uniquename=false,uniquelist=false,
            giveninits=true,hyperref=true]{biblatex}
\usepackage[colorlinks=true,linkcolor=blue,citecolor=blue,urlcolor=blue]{hyperref}
\usepackage[capitalize,nameinlink]{cleveref}

\makeatletter
\let\abx@macro@citeOrig\abx@macro@cite
\renewbibmacro{cite}{\bibhyperref{\let\bibhyperref\relax\relax\abx@macro@citeOrig}}
\let\abx@macro@textciteOrig\abx@macro@textcite
\renewbibmacro{textcite}{\bibhyperref{\let\bibhyperref\relax\relax\abx@macro@textciteOrig}}
\makeatother
\usepackage{pdflscape}

\newcommand{\fn}[1]{\textcolor{blue}{[#1]}}

\graphicspath{{fig/}}
\setlist{nosep}

\theoremstyle{plain}
\newtheorem{theorem}{Theorem}[section]

\newtheorem{lemma}[theorem]{Lemma}

\newtheorem{assumption}{Assumption}[section]
\newtheorem{remark}{Remark}
\theoremstyle{definition}
\newtheorem{definition}{Definition}[section]
\newtheorem{problem}{Problem}[section]
\newtheorem{model}{Model}[section]

\crefname{problem}{Problem}{Problems}
\Crefname{problem}{Problem}{Problems}
\crefname{model}{Model}{Models}
\Crefname{model}{Model}{Models}
\crefname{application}{Application}{Applications}
\Crefname{application}{Application}{Applications}

\definecolor{Pcol}{RGB}{20,20,140}
\definecolor{Scol}{RGB}{140,20,20}
\definecolor{Qcol}{RGB}{20,90,45}
\definecolor{propcol}{RGB}{140,20,20}

\newcommand{\cQ}{\mathcal{Q}}
\newcommand{\cG}{\mathcal{G}}
\newcommand{\E}{\mathbb{E}}
\newcommand{\R}{\mathbb{R}}
\newcommand{\I}{\mathbb{I}}
\newcommand{\Wone}{W_1}
\newcommand{\ATE}{\mathrm{ATE}}

\newcommand{\MSE}{\mathrm{MSE}}

\DeclareMathOperator{\Var}{Var}
\DeclareMathOperator*{\argmin}{arg\,min}
\DeclareMathOperator*{\argmax}{arg\,max}

\newcommand{\tS}{\tilde{S}}
\newcommand{\tQ}{\tilde{Q}}
\newcommand{\tcQ}{\tilde{\cQ}}
\newcommand{\Smilp}{S^{*}_{\mathrm{MILP}}}
\newcommand{\Sdro}{S^{*}_{\mathrm{DRO}}}
\newcommand{\plan}{\pi}

\algrenewcommand\algorithmicrequire{\textbf{Input:}}
\algrenewcommand\algorithmicensure{\textbf{Output:}}
\newcommand{\Ensures}{\Statex \hspace{\algorithmicindent}\textbf{Returns:} }

\newcommand{\prop}[2][]{{\color{propcol}\textsf{\scriptsize PROPOSED\ifx\\#1\\\else\ (#1)\fi}\ #2}}
\newcommand{\cut}[1]{{\color{gray}\sout{#1}}}

\title{ Experimental Site Selection Under \\Directional Distribution Shifts}
\author{Adam Bouyamourn\thanks{Assistant Professor and Faculty Fellow, Center for Data Science, New York University. }}
\date{\today}

\begin{document}

\maketitle

\begin{abstract}

Multi-site studies are widely used across the social sciences, education, and medicine to learn the effects of interventions across heterogeneous populations. But deployment populations, on which a given intervention is implemented, may differ systematically from the experimental population on which the intervention is evaluated. I propose a method for selecting sites with minimax guarantees against directional distribution shifts, which allows the researcher to guard against specified threats to the external validity of a multi-site study. I use Wasserstein Distributionally Robust Optimization to select sites with guarantees against directional shifts, provide a novel cutting-plane algorithm to implement the approach, and prove its optimality with respect to minimizing worst-case shifts. I highlight the empirical performance of the method with with two semi-synthetic simulations: I re-analyse the Tennessee STAR experiment, under the assumption that the study population underrepresented low-SES schools, and \citet{Crepon_2015}, which assesses the impact of increasing access to microcredit in rural Morocco, under the assumption that the study population was skewed towards villages with larger market potential. In these examples, I show that the WDRO method minimizes mean bias, average mean squared error, and $W_1$ distance under specified directional distribution shifts. 
\end{abstract}

\section{Introduction}\label{sec:intro}

Multi-site studies run a series of experiments across several locations or jurisdictions, with the goal of estimating treatment effects across experimental sites. They are widely used in the social sciences \citep{dunning2019information, blair2021community, banerjee2015multifaceted}, medicine \citep{rossouw2002risks, who2021repurposed}, and education \citep{Raudenbush2020, Orr2019}. 

They are a natural way to evaluate public policy interventions, which are often implemented at the level of geographic or institutional clusters, such as schools, hospitals, and legal jurisdictions. The effects of those interventions vary with the characteristics of the clusters, and with the composition of the subjects in them \citep{Bloom2017, Raudenbush2020}. 
Multi-site studies are popular because they have a suggestive ability to provide evidence about the \textit{scope} of an empirical finding: whether an intervention produces results across many jurisdictions, whether it replicates, and whether its effects generalize or depend on context \citep{dunning2019information, Bloom2017}.

In practice, conclusions drawn from a given experimental sample may not carry over to the population of interest. When deployment populations on which a policy is enacted differ from the population on which an intervention was evaluated, estimates derived from the experimental sample may not accurately represent treatment effects in the target population, undermining the external validity and practical utility of research findings \citep{shadish2002experimental, Findley2021,EGAMI_HARTMAN_2023, pearl_bareinboim_2011, Pearl_2014, Rudolph2021, Egami2021, rudolph2024improvingefficiencytransportingaverage}. This can be described as a problem of \textit{distribution shift} from an observed experimental population to a deployment population of interest.

Distribution shift can take on a number of concrete forms. Characteristics and the composition of populations may change over time \citep{bansak2023learningrandomdistributionalshifts, rothenhausler2023distributionallyrobustgeneralizableinference, Saville2022}, subgroups of particular interest may be underrepresented \citep{Tan2022, hu2024minimaxregretsampleselectionrandomized}, criteria for study inclusion may systematically misrepresent the population \citep{Olsen2016, Olsen2022, Allcott_2015}, or data used to plan studies may be subject to measurement error \citep{Bound_2001}.  Composition of available experimental sites may differ systematically from the policy-relevant population on which the researcher wishes to experiment \citep{Allcott_2015}, or minority groups may be systematically underrepresented in selected experimental units \citep{Tan2022,hu2024minimaxregretsampleselectionrandomized}. 

The forms of distribution shift to which a given study may be subject are often well-known to researchers in an applied area. 
In political science and economics, differences in institutional quality \citep{BOLD_2018}, trust in institutions \citep{Cheeseman_2023}, rural-urban mix \citep{dehejia2019localglobalexternalvalidity}, and racial and ethnic context \citep{Anoll_Davenport_2024, Hassell_2021}, can each be the source of substantive differences in the transportability of conclusions from one context to another. 

In education, school districts enrolled in an experiment may be unrepresentative of the distribution of schools, leading to overly optimistic assessments of policy impacts \citep{Olsen2016, Olsen2022}. A specific controversy centers on the evaluation of the benefits of charter schools, in which the marginal effect of charter school attendance is compared between students who won a lottery to attend an oversubscribed charter school, and students who did not win the lottery, and instead attended a public school \citep{Cohodes2024}. These studies can conceal variation within the population of charter schools, however, by censoring the distribution of charter schools \citep{Angrist2013}. 

In medicine, elderly patients and patients with common medical conditions are frequently excluded from randomized control trials  \citep{VanSpall2007, Zulman2011}, while cancer trials systematicaly underenroll minority groups, among whom specific varieties of cancer are likely to be more prevalent \citep{Oh2015}. Women are also often frequently underenrolled in clinical trials \citep{Daitch2022}. These exclusions can lead to overly optimistic assessments of drug efficacy \citep{Eichler2011}, and of side effects in minority populations \citep{Ramamoorthy2015}. 

This is analogous to the problem of selection bias in an observational study, where units opt into treatment on the basis of their expected returns from treatment \citep{heckman_1974, munro2025treatmenteffectsmarketequilibrium}. A subtlety, however, is that treatment effect estimates are unbiased in the resulting multi-site trial, because a randomized experiment is conducted after the sites have been selected. The bias instead is in the gap between the experimental population and the population of interest, and cannot be diagnosed from the results of the experiment alone.  

I propose a method for selecting sites in a multi-site study with guarantees against worst-case directional shifts in the deployment population. This allows the user to specify ex ante a threat to the external validity of a study by selecting a specific direction that maps onto a substantive threat to a study's transportability. 

I first prove a new upper bound on the mean squared error of the population average treatment effect in the site selection problem, using optimal transport \citep{villani2008optimal, peyre2019computational}. The bound decomposes the bias of a selection into two parts: bias due to observed site-level moderators, and bias due to unobserved moderators under the optimal transport plan on the observed ones. This gives a feasible target for optimization: minimize the transport distance between the site-level moderators of the selection and those of a hypothetical deployment population.

I write the planner's problem as a minimax problem. The planner chooses experimental sites to minimize the error of the selection against a set of adversarial distributions, generated by a model of distribution shift. I provide two parametric models of distribution shift: \textit{secular drift}, which is modelled as translation with respect to a direction $\theta$, and \textit{selection bias}, which is modelled as an exponential tilt with weights that depend on $\theta$. 

I then use Wasserstein Distributionally Robust Optimization to choose a site selection with worst-case guarantees against the form of shift in the specified direction. I develop a novel cutting-plane algorithm that solves a minimax version of the site selection problem, and prove that it terminates with optimality guarantees on the resulting selection. 

I reanalyse the Tennessee STAR experiment, and \citet{Crepon_2015}, a rural micocredit study in Morocco, under the assumption that their designs faced specific threats to external validity: in the former, that schools included in the study population were higher SES than schools in which the intervention would have been deployed; in the latter, that villages included in the study population had greater economic potential than villages that the policy would be deployed on. I show that the Wasserstein DRO method performs favorably in terms of mean bias and worst-case MSE under distribution shift compared to existing methods, which do not explicitly model the existence of adversarial distribution shifts. 

I conclude with guidance on how to specify a direction and radius of robustness in practice, the need to select site-level covariates, and a suggestion to focus more resources on gathering information at the experimental planning stage about site-level features, and the composition of policy-relevant groups.

\subsection{Related Literature}\label{sec:related}

Recent work in site selection proposes methods to choose experimental locations that are designed with external validity in mind \citep{gechter2024selecting, Egami_Dainlee_2024, olea2024externallyvalidselectionexperimental}. 

\citet{Egami_Dainlee_2024} introduced explicit optimization methods for site selection in political methodology, and contributed to defining the problem of site selection. Their approach, based on the synthetic control method, uses optimization to select included sites that closely approximate sites that are not included in the selection, by estimating balancing weights \citep{Abadie_2003, abadie_2010, abadie2025}. Their goal is to have a high-quality weighted average representation of non-selected sites. In practice, this can be thought of as ensuring that non-selected sites are within the convex hull of selected sites. 

Their formal result is an upper bound on the worst-case mean squared error of the predicted site-level effects under a smoothness condition on the effect surface. Given a target population of $N$ sites with covariates, synthetic purposive sampling selects $K$ sites and weights so that each non-selected site is approximated by a weighted average of the selected sites, in the manner of the synthetic control method \citep{Abadie_2003, abadie_2010, abadie2025}, with a penalty on the use of outlying sites. Its guarantee is an upper bound on mean squared error of the predicted site-level effects under a smoothness condition on the effect surface. 

A key difference is that their solution set intends to solve for a convex hull that contains the PATE. This can lead to solution sets that include outliers. Their default setting is a regularize convex hull, which shrinks the convex hull by penalizing inclusion of outlying sites. This approach is motivated by practical concerns, but does not offer a formal guarantee on robustness to hypothetical deployment contexts. 

\citet{Tipton_2013} develops a subclassification estimator that reweights an existing experimental sample toward a population on strata of an estimated selection score; \citet{Tipton2013b} turn the same idea into a design, stratifying the population of sites on the score or on clusters formed from the covariates and selecting sites within strata so that the sample's composition matches the population's across strata. The guarantee is that of stratified sampling: the sample effect is unbiased for the population effect when the effect surface is constant within strata, and its bias is bounded by within-stratum variation otherwise. The population is fixed and observed.

\citet{olea2024externallyvalidselectionexperimental} is the closest method to the current paper. They formulate site selection as a minimax decision problem: a policymaker chooses, for each site in a known set of policy sites, whether to adopt a treatment, and the effect surface is Lipschitz with a known constant. They show that selecting sites by solving the $k$-median problem solves a minimax-regret problem. 

This paper draws on the \textbf{Distributionally Robust Optimization} (DRO) literature, a set of methods developed in operations research that find solution sets with guarantees against worst-case performance within the radius of a given solution \citep{Ben-Tal_2013,esfahani2017datadrivendistributionallyrobustoptimization, kuhn2024wassersteindistributionallyrobustoptimization, blanchet2021statisticalanalysiswassersteindistributionally, Blanchet_Murthy_2018, blanchet2024distributionallyrobustoptimizationrobust, duchi2020learningmodelsuniformperformance, levy2020, bertsimas2023distributionallyrobustcausalinference}. DRO methods approach the problem of distributional uncertainty by providing statistical guarantees that a given solution is robust to a worst-case shift of the data. These methods provide insurance against poor performance within a specified neighborhood of the empirical solution \citep{Luo2020, duchi2020learningmodelsuniformperformance}. 
In statistics and causal inference, distributionally robust methods have been used for estimation under covariate shift and sample-selection bias: \citet{duchi2020learningmodelsuniformperformance} for uniform performance over subpopulations, \citet{bertsimas2023distributionallyrobustcausalinference} for causal effects from observational data, and \citet{zhang2024minimaxregretestimationgeneralizing} for generalizing heterogeneous effects across sites.

Minimax criteria for statistical decisions originate with \citet{wald1950}. In treatment choice the relevant criterion is minimax regret \citep{Manski2004, Stoye2009, HiranoPorter2009}. In experimental design, \citet{Kallus2018} shows that the minimax-variance treatment assignment over a class of outcome functions is a balance criterion determined by that class, and \citet{Kallus2021} characterizes when
randomization is minimax; \citet{Bertsimas2015} optimize assignment directly; \citet{hu2024minimaxregretsampleselectionrandomized} choose how many units to draw from each subpopulation to minimize worst-case regret over effect surfaces. This paper is closely related to the work of \citet{kallus2020optimal} and \citet{Bertsimas2015}, but considers a problem in which the data-generating process is assumed to be subject to distribution shift. Randomization is not necessarily minimax for this problem, and optimization of the nominal problem is not optimal for the shifted problem. 

\Cref{sec:method} develops the site selection problem under distribution shift, states bounds on the quantity of interest, and describes two parametric models of distribution shift. \Cref{sec:algorithms} motivates a nominal optimization problem, the cutting-plane algorithm used to solve the distributionally robust optimization problem, and states theoretical guarantees on the algorithm. \Cref{sec:empirical} provides an assessment of extant site selection methods using data from the Tennessee STAR and \citep{Crepon_2015} microcredit experiments. \Cref{sec:conclusions} concludes.

\section{Theory and Methods}\label{sec:method}
\subsection{The Site Selection Problem Under Distribution Shift}\label{sec:problem}

\begin{figure}[H]
\begin{centering}
\begin{tikzpicture}[>={Stealth[length=2.6mm]}, line width=0.6pt]
  \draw[Pcol] (0,0)         ellipse [x radius=3.20, y radius=2.60];
  \draw[Qcol] (3.90,-0.45)  ellipse [x radius=2.90, y radius=2.40];
  \draw[Scol] (-1.05,0.25)  ellipse [x radius=1.94, y radius=1.58];
 
  \node[Pcol] at (-2.56,-2.08) {$P$};
  \node[Qcol] at ( 6.22,-2.37) {$Q$};
  \node[Scol] at (-1.05, 2.10) {$S$};
 
  \fill[Scol] (-1.05,0.75) circle (1.9pt);
  \node[Scol, anchor=south, inner sep=3pt] at (-1.05,0.81) {\small $\widehat{\ATE}(S)$};
 
  \coordinate (patep) at (-1.05,-0.30);
  \fill[Pcol] (patep) circle (1.9pt);
  \node[Pcol, anchor=north, inner sep=3pt] at (-1.05,-0.36) {\small $\ATE(P)$};
 
  \coordinate (pateq) at (3.90,-0.45);
  \fill[Qcol] (pateq) circle (1.9pt);
  \node[Qcol, anchor=north west, inner sep=3pt] at (4.00,-0.55) {\small $\ATE(Q)$};
 
  \draw[->, shorten <=4pt, shorten >=4pt] (patep) -- (pateq);
\end{tikzpicture}

  \end{centering}
\end{figure}

Let $P$ be a set of candidate sites observed during the planning stage. The experimental planner's goal is to select a subset of sites $S \subset P$, where the subset is of size $K$, in order to minimize a realized loss on the deployed population of interest. 

Each site $s$ has associated covariates $(X, U)$, where $X \in \R^d$ are observed at the planning stage, and $U \in \R^k$ are not observed. $X$ is comprised of site-level covariates $X_s$, and aggregates of individual-level covariates $\bar{X}_i$, so that $X = \{ X_s \cup \bar{X}_i\}$. We refer to these collectively as \textit{site characteristics}, or  site-level covariates, and do not distinguish between these elements in the sequel. Let $P_X$ be the empirical distribution of observed site characteristics, $P_U$ the distribution of unobserved site covariates, and $P_{X,U}$ their joint distribution. Likewise $S_X$, $S_U$ and $S_{X,U}$ refer to their respective sample distributions.

We suppose that the policy being evaluated is enacted on a deployment population $Q$. We take $Q$ to belong to a set of distributions $\cQ_\rho$ generated from $P_X$ by specific transformations corresponding to particular models of distribution shift (see \Cref{sec:varieties}). We suppose that these distributions are within a radius $\rho$ of the original distribution $P_X$.  

First, we define the Average Treatment Effect, as a function of an arbitrary distribution $F$ which we take as the primary estimand of interest to the planner in a multi-site study. 

\begin{definition}[Average Treatment Effect]\label{def:ate}
Let $\tau: \R^d \times \R^k \to \R$ be the treatment effect as a function of covariates. Then, for a generic joint distribution $F$ such that $(X, U) \sim F$,
\begin{equation*}\label{eq:pate}
ATE(F) = \int \tau(x, u) \, dF(x, u) .
\end{equation*}
\end{definition}
When F = P, this is the Population Average Treatment Effect, and when F = Q, this is the Average Treatment Effect in the Deployment Population. To avoid the ambiguity involved in `PATE' under distribution shift, I prefer the more specific $ATE(P)$ or $ATE(Q)$ below. 
\noindent We define the \textbf{deployment error}, with respect to a deployment population $Q$:

\begin{definition}[Deployment error]\label{def:deployment_error}
The deployment error of a selection $S$ with respect to a deployment population $Q$ is:
\begin{equation*}\label{eq:mse}
\MSE_Q(S) = \E_{Z \sim \Psi, W\sim \Phi}\big[(\ATE(Q) - \widehat{\ATE(}S))^2\big].
\end{equation*}
\end{definition}

Where the expectation is taken over the randomness in unit-level treatment assignment $Z$ according to a design $\Psi$, and unit-level experimental inclusion $W$ according to a sampling process $\Phi$, and $\widehat{\ATE}(S)$ is the realized estimate of the ATE generated from running an experiment on the subset $S$. 

In this paper, we are interested in minimizing the \textbf{worst-case deployment error} with respect to a set of distributions that are within a radius $\rho$ of our observed distribution: 

\begin{definition}[Worst-case deployment error]\label{def:deployment_error}
The worst-case deployment error of a selection $S$ with respect to a deployment population $Q$ within radius $\rho$ of $P_X$ is:
\begin{equation*}\label{eq:mse}
\sup_{Q \in {\cQ_\rho}} \; \MSE_Q(S) = \sup_{Q \in {\cQ_\rho}} \; \E\big[(\ATE(Q) - \widehat{\ATE(}S))^2\big].
\end{equation*}
\end{definition}

The Researcher's goal is to minimize the worst-case deployment by choice of a subset $S$:

\begin{problem}[Site selection under Distribution Shift]\label{prob:robust}
Given candidate sites $P$, a budget $K$, and a set of deployment populations $\cQ_\rho$, the Researcher's problem is:
\begin{equation*}\label{eq:robust_problem}
\min_{S \subset P,\; |S| = K} \; \sup_{Q \in {\cQ_\rho}} \MSE_Q(S).
\end{equation*}
\end{problem}

\noindent The inner supremum is
the deployment population that maximizes the deployment error of that choice, while the outer minimization describes the Researcher's choice of experimental sites.\footnote{In the setting without distribution shift, $\rho = 0$, $\cQ = \{P_X\}$ and \Cref{prob:robust} is $\min_{S} \MSE_{P_X}(S)$.}

\begin{figure}[htbp]

\begin{tcolorbox}[title={The Site Selection Problem Under Distribution Shift}]

\begin{enumerate}[leftmargin=*]\item \textbf{Planning.} The researcher defines the population of sites $P$ and observes site-level covariates $X$ for each candidate site. Given a budget constraint $K$, the researcher specifies a set of distribution shifts $\cQ_\rho$ to guard against, and chooses a subset $S$.  \item \textbf{Experimentation.} An experiment is run in each selected site. Individual-level covariates are observed in the selected sites. Treatment is assigned within sites. \item \textbf{Analysis.} The analyst estimates $\widehat{\ATE}(S)$, which also incorporates individual-level covariates, and may include with the declared design or welfare weights. \item \textbf{Real-world Deployment.} A covariate distribution $Q$, and the deployment error, $MSE_Q(S)$, are realized.\end{enumerate}

\end{tcolorbox}

\caption{The Researcher's Site Selection Problem}

\label{fig:sequence}

\end{figure}

\subsection{Assumptions}
\Cref{prob:robust} is infeasible without further assumptions on the relationship between the treatment effect function and covariates. We require that the treatment effect function is Lipschitz continuous in covariates (Lipschitz Continuity of $\tau$), and that there is one treatment effect function common across units. 
   
    \begin{assumption}[Lipschitz Continuity of $\tau$]\label{asm_Lipschitz}
The treatment effect function $\tau: \mathbb{R}^d\times \mathbb{R}^k \rightarrow \mathbb{R}$ is Lipschitz continuous with constant $L$: 
$$|\tau(x, u) - \tau(x', u')| \leq L (\|x - x'\| + \|u - u'\|) \quad \forall x, x' \in \mathbb{R}^d, \forall u, u' \in \mathbb{R}^k$$
    \end{assumption}
This ensures that treatment effects vary smoothly with covariates. When covariate values change, treatment effects must vary within an envelope defined by the size of the change of covariate values. This assumption is important, because it allows us to move from claims about covariates to claims about treatment effects.

We also require that the randomization design $\Psi$ is independent of the optimal choice of subset: 

\begin{assumption}[Independence of Experimental Design and Site Selection]\label{asm_indep}
The randomization and unit sampling distributions in any given site $\Psi_s, \Phi_s$ do not depend on the choice of experimental subset.
  \begin{equation*}
        Z_s | S = A \sim \Psi_s \textrm{ and } W_s | S = A \sim \Phi_s, \: \forall A \subseteq P \textrm{ such that } s \in A.  
  \end{equation*}
\end{assumption}

Two useful properties follow from \Cref{asm_indep}.

\begin{lemma}[Unbiasedness of site-level estimation]\label{lem:unbiased}
\begin{equation*}
    \E_{Z \sim \Psi, W \sim \Phi}[\widehat{\ATE}(S)] = \int\tau(x, u) \:dS_{X,U}(x,u) = \ATE(S)
\end{equation*} 
\end{lemma}

Define the site-specific estimation error $\Var(\widehat{\ATE}(S)) = \sigma^2_S$. Then:

\begin{lemma}[Bias--variance decomposition]\label{lem:biasvar}
\begin{equation*} 
\MSE_Q(S) = \big(\ATE(Q) - \ATE(S)\big)^2 + \sigma^2_S
\end{equation*}
\end{lemma}

The first term is controlled by choice of subset $S$, while the second term is controlled by choice of randomization design. 

\subsection{Bounds on the deployment error of a site selection}

The assumptions above allow us to move from statements about covariates to statements about the error on the deployment distribution. 

\subsubsection{Technical Preliminaries} 

First, it is helpful to define the notion of a \textit{transport plan}, which is a matrix that specifies a routing of mass from one distribution to another. 

\begin{definition}[Transport plan]
A \textbf{transport plan} between discrete distributions
$P_X = \sum_{i=1}^n p_i \delta_{x_i}$ and $Q_Y = \sum_{j=1}^m r_j \delta_{y_j}$
is a matrix $\{\pi_{ij}\}_{(i=1, j=1)}^{(n,m)}$ with $\pi_{ij} \geq 0$ such that
$\sum_{j=1}^m \pi_{ij} = p_i$ for each $i$ and $\sum_{i=1}^n \pi_{ij} = r_j$ for each $j$.
\end{definition}

Where $\delta$ is the Dirac delta. Row $i$ of $\pi$ sums $p_i$, the mass at $P_{x_i}$,  while column $j$ of $\pi$ sums to $r_j$, the mass at $Q_{x_j}$. 

In order to evaluate different transport plans, we need a way to assess the costs of a given proposed transport plan. A \textbf{cost function} describes the cost of travelling from $X$ to $Y$. We use the Euclidean distance as our cost function, so that
$c(X,Y) = \|X - Y\|_2 = \big(\sum_{l=1}^{d} (X_l - Y_l)^2\big)^{1/2}$.

The \textbf{optimal transport plan} is the plan $\pi^*$ that in fact minimizes the distance between $P$ and $Q$, for a given cost function $c(X, Y)$. That is,

\begin{definition}[Optimal Transport Plan]
A transport plan $\pi^*$ is optimal if 
$$\pi^* = \arg \underset{\pi}{\inf} \sum_{i=1}^n \sum_{j=1}^m \pi_{ij}||x_i - y_j||$$ 
\end{definition}

That is, if $\pi^*$ minimizes the cost of transporting mass from $P$ to $Q$.

We can think of the solution to the optimal transport as being the shortest possible distance between $X$ and $Y$, given the distributions P and Q.

Next, we define the 1-Wasserstein distance: 

\begin{definition}
For two distributions $\mu$ and $\nu$ on $\R^d$ with finite first moments, the 1-Wasserstein distance is
\begin{equation*}\label{eq:w1}
\Wone(\mu, \nu) = \inf_{\gamma \in \Gamma(\mu, \nu)} \int \|x - x'\| \, d\gamma(x, x'),
\end{equation*}
where $\Gamma(\mu, \nu)$ is the set of couplings of $\mu$ and $\nu$. For empirical distributions $P$ and $Q$, the coupling $\gamma$ is a \textit{transport plan} \citep{Kantorovich2006}. $\gamma^*$, the transport plan that distributes mass from P to Q at infimal cost, is called the \textit{optimal transport plan} \citep{villani2008optimal}.

\end{definition}

Notice that the definition of $W_1$ distance is stated in terms of the optimal transport plan. Indeed, the $W_1$ distance from $P \to Q$ formalizes the notion of the the shortest distance from $P \to Q$ in the $\ell-1$ norm: there are many ways to transport mass from P to Q, and $W_1$ is the distance associated with the lowest-cost way to do so. 

\subsubsection{Bounds on the deployment error}
It is first helpful to consider the case where $Q$ is known. In this case, we can derive the following upper bound on the deployment error:

\begin{theorem}[Bound on the deployment error when $Q$ is known and covariates are observed]\label{thm:oracle_Q}
Suppose \Cref{asm_Lipschitz} and \Cref{asm_indep} hold. Then, when all covariates are observed, for any selection $S$ and a known deployment distribution $Q$,
\begin{equation*}\label{eq:oracle}
\MSE_Q(S) \le L^2 \, \Wone(Q, S_X)^2 + \sigma^2_S ,
\end{equation*}
\end{theorem}

The bound is infeasible as a design criterion, because $Q$ is not observed when $S$ is chosen. Indeed, our goal is to guard against the worst-case $Q$ from a family of possible, unobserved shifts applied to our observed $P_X$.  

To state a feasible bound, we want to decompose the distance between $S$ and $Q$  into observable and unobservable components. Recall that an optimal transport plan is a matrix that specifies the lowest-cost routing of mass from one distribution onto another. Let $\gamma^*$ denote such a transport plan from $S_X$ to $Q_X$. We consider the difference between the unobservable \textit{conditional distributions} $S_{U|X}$, $Q_{U|X}$, under the optimal transport plan that maps $S_X$ onto $Q_X$:

\begin{definition}[Unobserved heterogeneity]\label{def:eta}
\begin{equation*}
\eta(S_U, Q_U) = \int \Wone\big( S_{U \mid X = x}, \, Q_{U \mid X = x'} \big) \, d\gamma^{*}(x, x') .
\end{equation*}
\end{definition}

This averages over the differences in unobservables across the selected set $S$ and the deployment population $Q$. It is a measure of the sensitivity of a selection $S$ to unobserved site-level heterogeneity in the deployment population $Q$. 

We can then plug this quantity into our bound: 

\begin{theorem}[Bound on the deployment error with unobserved moderators]\label{thm:eta_Q}
Let $\gamma^{*}$ be an optimal coupling for $(S_X, Q_X)$. Then, under \Cref{asm_Lipschitz} and \Cref{asm_indep}
\begin{align*}\label{eq:eta_bound}
\MSE_Q(S) \le L^2 \big[ \Wone(S_X, Q_X) + \eta(S_U, Q_U) \big]^2 + \sigma^2_S,
\end{align*}
\end{theorem}

This bound decomposes the deployment error into an observable and unobservable component. We can therefore choose $S$ to minimize the observable component of the bound, giving us the feasible problem:

\begin{problem}
    \begin{equation*}\label{eq:robust_problem}
\min_{S \subset P,\; |S| = K} \; \sup_{Q \in {\cQ_\rho}} W_1(S_X, Q_X).
\end{equation*}
\end{problem}

$\cQ_\rho$ is a set of distributions constructed from the observed distribution $P_X$. We next explain how these adversarial distributions are generated.  
\subsection{Directional Distribution Shift}\label{sec:varieties}

We restrict our focus to specific varieties of distribution shift that are likely to arise in practice in applied policymaking contexts. In particular, we restrict our attention to \textit{directional distribution shifts}.

\begin{definition}[Directional distribution shift family]\label{def:family} Let $\mathscr{T} = \{T_\theta : \theta \in \Theta\}$ be a class of transformations from $\mathcal{P}(\R^d) \to \mathcal{P}(\R^d)$, with $Q_\theta = T_\theta(P_X)$, a distribution induced by applying a specific transformation $T_\theta$ to the empirical distribution $P_X$. Define a restricted parameter space $\Theta_\rho = \{\theta \in \Theta: W_1(Q_\theta, P_X) \leq \rho \}$, that is, the set of values of $\theta$ for which the induced distribution is within a 1-Wasserstein radius of $\rho$ of the original empirical distribution. Then, a distribution shift family of budget $\rho$ indexed by parameter $\theta \in \Theta_\rho$ is a family of distributions: 
\begin{equation*}
\cQ_\rho(\Theta) = \{ T_\theta(P_X) : \theta \in \Theta_\rho, W_1(T_\theta(P_X), P_X) \leq \rho \} 
\end{equation*}
\end{definition}

\begin{remark}
Each $\theta \in \Theta$ is a vector in $\R^m$, and its direction $\theta / \lVert \theta \rVert$ indexes a direction of shift. The set $\Theta$ therefore encodes which shifts the analyst regards as possible.
\end{remark}

When the set of possible deployment populations belongs to a directional shift family $\cQ_\rho(\Theta)$, the problem becomes:

\begin{problem}[DRO problem over a directional distribution shift family]\label{restricted_problem}
\begin{equation*}
    \min_{S : |S| = K} \; \sup_{Q_\theta \in \cQ_\rho(\Theta)} \Wone(Q_\theta, S_X)
    \end{equation*}
\end{problem}

Where $Q_\theta$ now belongs to a directional distribution shift family, and the supremum picks out the worst-possible shift belonging to such a family. Choosing specific distribution shift families gives us robustness against particular models of distribution shift. These different versions of distribution shift have different substantive policy consequences, and we allow the analyst to decide which kind is relevant to their given their research goals and the application.

\begin{model}
    [Secular drift (Translation)]\label{mod:translation}
Let $\Theta \subseteq \R^d$ be a set of possible directions along which the covariates may drift, and let $T_\theta$ shift every site by the vector $\theta$:
\begin{equation*}
Q_\theta = \frac{1}{|P|} \sum_{s=1}^{|P|} \delta_{X_s + \theta} .
\end{equation*}
\end{model}

For instance, deployment sites could be systematically poorer, more rural, or larger than the planning data, by up to $\rho$ standard deviations.

\noindent 
\begin{model}[Selection bias (Exponential tilting)]\label{mod:tilt}
Let $\Theta \subseteq \R^d$ be a set of directions along which sites may be over- or under-represented, and let $T_\theta$ reweight the sites log-linearly:
\begin{equation*}
Q_\theta = \sum_{s=1}^{|P|} w_s(\theta) \, \delta_{X_s} ,
\qquad
w_s(\theta) = \frac{\exp \{ \theta^\top X_s\}}
{\sum_{r=1}^{|P|} \exp \{\theta^\top X_r\}} .
\end{equation*}
\end{model}

This model of distribution shift is applied in \citet{tibshirani2020conformalpredictioncovariateshift}, and mirrors the problem of site selection bias described in \citet{Allcott_2015}. Suppose that sites on the candidate list were non-randomly sampled from a deployment population, so that each site $s$ had sampling weight $\psi_s \neq \frac{1}{|P|}$. In order to reweight the candidate population $P_X$ back to the original deployment population $Q$, we would want to solve for Horvitz-Thompson weights $q_s \propto \frac{|P|^{-1}}{\psi_s}$. If we suppose that the weights $\psi_s \propto \exp \{-\theta^\top X_s\}$, we get the exponential tilting model above.

\section{Algorithms and Computation}\label{sec:algorithms}

\subsection{Warm-up: Optimal Transport and the Site Selection Problem}\label{sec:warm-up}

 To motivate the distributionally-robust optimization problem used to solve the site selection problem, it is helpful first to consider the simpler optimization problem when the deployment population is the observed population $P_X$.
 
\begin{problem}[Optimization problem over an observed distribution]\label{restricted_problem}
\begin{equation*}
    \min_{S : |S| = K} \; \Wone(P_X, S_X)
    \end{equation*}
\end{problem}

It is helpful to substitute into this problem the definition of the discrete $1$-Wasserstein distance, to get the following problem statement:

\begin{problem}[Optimization problem over an observed distribution]\label{prob:nominal}
\begin{equation*}
    \min_{S \,:\, |S| = K} \;\; \min_{\gamma \in \Gamma(P_X, S_X)} \;\;
    \sum_{j=1}^{|P|} \sum_{k \in S} \gamma_{jk} \, \lVert X^P_j - X^S_k \rVert ,
\end{equation*}
\end{problem}
This problem has an inner minimization step, and an outer minimization step. These are solved simultaneously as a Mixed Integer Linear Program, but it is helpful to provide intuition for both steps. 

The inner minimization step is an optimal transport problem. It takes a fixed selection of sites $\tilde{S}$, and finds an optimal transport plan $\gamma^*(\tilde{S})$, that minimizes the distance between $P_X$ and $\tilde{S}_X$, given $\tilde{S}$ fixed. Because it is an optimal transport problem, it can be stated as a linear program \citep{Kantorovich2006, Dantzig51}. 

The outer minimization step chooses the set $S^*$ that minimizes the total objective, given the optimal transport plan $\gamma^*(S)$ for each possible subset $S$. This step adds integer constraints, because site selection is a discrete selection problem. Selecting sites given fixed losses is a Mixed Integer Problem. The combined problem is therefore a Mixed Integer Linear Program. 

The combined problem is solved by branch-and-bound \citep{LandDoig1960, Dakin1965}. First, the integer constraints are relaxed, reducing the MILP to a Linear Program with fractional site assignments. Then, the solver evaluates the objective for each pair of subproblems, with the fractional assignments fixed to integer constraints. A subproblem is discarded when its relaxation is dominated by the current best integer solution. 

The solution set $S_{\textrm{MILP}}^*$ generated from this procedure minimizes the observed distance between $S_X$ and $P_X$, which is the case when distribution shift is assumed not to occur, the deployment population is the observed population, and $\rho = 0$. 

\begin{figure}
\begin{tcolorbox}
For a target distribution $P_X$ and selection distribution $S_X$:
\begin{align*}
\min_{s, \pi} \quad & \sum_{j=1}^{|P|} \sum_{k=1}^{|P|} \pi_{jk} \, \|X_j^P - X_k^S\| \\
\text{subject to:} \quad & \\
& \sum_{k=1}^{|P|} s_k = K \tag{Site budget} \\
& \sum_{k=1}^{|P|} \pi_{jk} = \frac{1}{|P|} \quad \forall j \tag{Target marginal} \\
& \sum_{j=1}^{|P|} \pi_{jk} = \frac{s_k}{K} \quad \forall k \in P \tag{Selection marginal} \\
& \pi_{jk} \geq 0 \quad \forall j,k \in P \tag{Nonnegativity} \\
& s_k \in \{0,1\} \quad \forall k \in P \tag{Binary selection}
\end{align*}
\end{tcolorbox}
\caption{MILP formulation of the Site Selection Problem}
\label{MILP_box}
\end{figure}

\subsection{Cutting-plane algorithm for site selection under distribution shift}

Now recall our optimization problem of interest: 

\begin{problem}[DRO problem over a directional distribution shift family]\label{restricted_problem}
\begin{equation*}
    \min_{S : |S| = K} \; \sup_{Q_\theta \in \cQ_\rho(\Theta)} \Wone(Q_\theta, S_X)
    \end{equation*}
\end{problem}

It is instructive to note that this is a minimax problem, and we solve it by implementing a cutting-plane algorithm. 

\begin{algorithm}[p]
\caption{Cutting-plane method for site selection under distribution shift}
\label{alg:formal}
\begin{algorithmic}[1]
\Require Candidate covariates $X^P_1, \ldots, X^P_{|P|}$; budget $K$; shift family $\cQ_\rho(\Theta)$; mesh $h$; nominal selection $\Smilp$; tolerance $\epsilon \ge 0$
\Ensure Robust selection $\Sdro$
\Statex
\Statex \hrulefill
\Procedure{EvaluateSiteSelectionUnderQ}{$S$, $Q$} 
  \Ensures $\Wone(S, Q)$, the cost of an optimal transport plan from $S \to Q$
  \State $C_{js} \gets \lVert X^Q_j - X^P_s \rVert$ \ for $j \le |P|$, $s \in S$
  \State $\tilde\plan \gets \argmin \Big\{ \sum_{j}\sum_{s \in S} \plan_{js} C_{js} \ :\ \plan \ge 0,\ \ \sum_{s \in S} \plan_{js} = w_j\ \forall j,\ \ \sum_{j} \plan_{js} = \tfrac{1}{K}\ \forall s \in S \Big\}$
  \State \Return $\sum_{j,s} \tilde\plan_{js} C_{js}$
\EndProcedure
\Statex
\Procedure{AdversaryBestResponse}{$\tS$}
  \Ensures Highest cost adversarial shift $\tilde{Q}$ over a grid, given a site selection $\tilde{S}$
  \For{$Q \in \cQ_h$}
    \State $\Wone(Q, \tS_X) \gets \Call{EvaluateSiteSelectionUnderQ}{\tS, Q}$
  \EndFor
  \State $\tQ \gets \argmax_{Q \in \cQ_h} \Wone(Q, \tS_X)$
  \State \Return $\tQ$
\EndProcedure
\Statex
\Procedure{PlannerBestResponse}{$\tcQ$}
  \Ensures A selection $\tilde{S}$ that minimizes the error induced by the worst shift $\tilde{Q}$ in a collection $\tilde{\cQ}$ via MILP
  \State $C^{(Q)}_{js} \gets \lVert X^Q_j - X^P_s \rVert$ \ for $Q \in \tcQ$, $j \le |P|$, $s \le |P|$
  \State $\begin{aligned}[t]
    \tS \gets \argmin_{\I\{s \in \tS\} \in \{0,1\},\ \gamma^{(Q)} \ge 0,\ t \in \R}\ & t \\[-1pt]
    \text{subject to}\quad & t \ \ge\ \textstyle\sum_{j,s} \gamma^{(Q)}_{js} C^{(Q)}_{js} && \forall Q \in \tcQ, \\[-1pt]
    & \textstyle\sum_{s} \I\{s \in \tS\} = K, \\[-1pt]
    & \textstyle\sum_{s} \gamma^{(Q)}_{js} = w^{(Q)}_j && \forall j,\ \forall Q \in \tcQ, \\[-1pt]
    & \textstyle\sum_{j} \gamma^{(Q)}_{js} = \I\{s \in \tS\}/K && \forall s,\ \forall Q \in \tcQ
  \end{aligned}$ 
  \State \Return $\tS$
\EndProcedure
\Statex \hrulefill
\Statex
\State $\cQ_h \gets \{ Q_\theta : \theta \in \cG_h \}$, where $\cG_h \subset \Theta_\rho$ is a grid of mesh $h$ with $0 \in \cG_h$ 
\State $\tcQ^{(0)} \gets \{P_X\}$; \quad $\tS^{(0)} \gets \Smilp$; \quad $t \gets 0$
\Repeat
  \State $\tQ^{(t+1)} \gets \Call{AdversaryBestResponse}{\tS^{(t)}}$ 
  \State $\tcQ^{(t+1)} \gets \tcQ^{(t)} \cup \{\tQ^{(t+1)}\}$ \Comment{Worst-case shift added to collection}
  \State $\tS^{(t+1)} \gets \Call{PlannerBestResponse}{\tcQ^{(t+1)}}$ 
  \State $\mathrm{UB}^{(t)} \gets \Call{EvaluateSiteSelectionUnderQ}{\tS^{(t)}, \tQ^{(t+1)}}$ 
  \State $\mathrm{LB}^{(t+1)} \gets \max_{Q \in \tcQ^{(t+1)}} \Call{EvaluateSiteSelectionUnderQ}{\tS^{(t+1)}, Q}$ 
  \State $t \gets t + 1$
\Until{$\min_{u < t} \mathrm{UB}^{(u)} - \mathrm{LB}^{(t)} \le \epsilon$}
\State \Return $\Sdro \gets \tS^{(u^{\star})}$ with $u^{\star} \in \argmin_{u < t} \mathrm{UB}^{(u)}$
\end{algorithmic}
\end{algorithm}
This problem requires us to generate a set of adversarial distribution shifts in observed covariates according to the model of distribution shift chosen by the analyst:
\begin{equation*}
Q_{\theta^{(1)}}, Q_{\theta^{(2)}}, \cdots \in \cQ_\rho(\Theta)
\end{equation*}

We specify a mesh of width $h$ over the parameter grid $\mathcal{G}_h \subset \Theta$. This generates a grid of parameter values $\mathcal{G}_h = \theta_1, \dots, \theta_{|\mathcal{G}_h|}$, and we can generate one distribution $Q$ for each point on this grid. We refer to this grid of shifted distributions as $\cQ_h$. 

The cutting-plane algorithm requires three subroutines. The first, \textsc{EvaluateSiteSelectionUnderQ}$(S,Q)$, takes a proposed site selection $\tilde{S}$, and finds an optimal transport plan between $\tilde{S}$ and $Q$. This is a linear program, because the site selection is fixed. We use this as a general utility to evaluate the cost of different adversarial distributions $Q$ with respect to a proposed site selection $\tilde{S}$. ($\tilde{\cdot}$ is used to denote an intermediate selection of the procedure.)

The next is the adversary's move, \textsc{AdversaryBestResponse}$(\tilde{S})$. This takes a proposed site selection $\tilde{S}$ and evaluates it against every distribution shift generated along the grid $\cQ_h$. It yields a vector of costs $W_1(Q, \tilde{S})$, for each Q in $\cQ_h$. Taking the maximum cost, 

For any observed sites selection $\tilde{S}$, the adversary can evaluate the loss of $\tilde{S}$ over the whole grid of generated shifts, and return whichever shift $\tilde{Q}$ generated the largest observed optimal transport plan cost. 

It also requires us to solve a version of the MILP above for a set of observed adversarial distributions. This helper function is called \textsc{PlannerBestResponse}$(\tilde{Q)}$. In particular, we want to solve the MILP for the worst-case adversarial distribution. To do this, we specify the epigraph of the loss due to the site selection, $t$, and set it greater than or equal to the largest cost of the worst $\tilde{Q}$ observed in a specific set $\tilde{\cQ}$. This formalization trick allows us to minimize a linear program against the maximum loss, which is nonlinear. 

The algorithm then proceeds as follows. First, the algorithm is initialized using the solution set $S^*_{\textrm{MILP}}$ from \Cref{sec:warm-up}, and a grid of parameter values is generated. Second, the adversary evaluates the cost of a set of optimal transport plans from  $S^*_{\textrm{MILP}}$ to the set of adversarial distributions $\cQ_h$, and selects the highest-cost adversarial distribution $\tilde{Q}$. This worst-case distribution is added to a storage set of adversarial distributions $\tilde{\cQ}$. Third, the planner chooses a best response to the worst-case distribution in the set $\tilde{\cQ}$ via MILP. 

The value of the worst-cost transport plan from the previous selection, evaluated before the planner chooses a new set of sites, here, $S^*_{\textrm{MILP}}$, is stored as the upper bound (UB) on the loss: it is the worst-case loss of a feasible selection. The value of the worst-cost transport plan from the current selection is stored as the lower bound in the loss: the adversary can only increase the loss by re-selecting an adversarial distribution. 

We therefore have a nonincreasing minimum upper bound on the worst-case transport loss, and a nondecreasing lower bound on the worst-case transport loss. The algorithm terminates when $\underset{u < t}{\min}UB^{(u)} - LB^{(t)} \leq \epsilon$, at which point the worst-case transport loss over the parameter grid is within epsilon of its optimal value, and the site selection is output as $S^*_{\textrm{DRO}}$.  

\subsection{Guarantees}\label{sec:guarantees}

\begin{theorem}[Guarantee for \Cref{alg:formal}]\label{thm:alg}
Suppose $\Theta_\rho$ is bounded, $\cG_h \subset \Theta_\rho$ is a grid such that every
$\theta \in \Theta_\rho$ lies within $h$ of some grid point, and \textsc{PlannerBestResponse} returns
an exact minimizer. Then \Cref{alg:formal} terminates after at most $|\cQ_h|$ iterations, and its
output $\Sdro$ satisfies
\begin{equation*}\label{eq:guarantee}
\sup_{\theta \in \Theta_\rho} \Wone\big(Q_\theta, S^{*}_{\mathrm{DRO},X}\big)
\;\le\;
\min_{S : |S| = K} \; \sup_{\theta \in \Theta_\rho} \Wone\big(Q_\theta, S_X\big)
\;+\; \epsilon \;+\; c\,h ,
\end{equation*}
Where $c$ is a Lipschitz constant. 
\end{theorem}

\section{Empirical Studies}\label{sec:empirical}
\subsection{Tennessee STAR: Undersampling of low-SES schools}
\subsubsection*{Simulation Design}

The Tennessee STAR experiment studied the effect of smaller class sizes on learning outcomes at 75 schools. It found that smaller class sizes improved test scores \citep{Mosteller1995}. 

A plausible threat to validity would occur if schools enrolled in the STAR experiment were more socioeconomically advantaged than schools in which the intervention was eventually implemented.  

Because we observe outcomes and covariates for 75 schools, we can do two things. First, we can partition schools into a \textit{candidate set} and a \textit{deployment set}. The candidate set is observed by the methods, and used to make a decision about where to experiment. We can manipulate the deployment set according to a model of distribution shift in order to induce distribution shift in the held-out sites. 

We first project covariates into principal component space. The first principal component loads on the covariates \textit{free-lunch share}, \textit{minority share}, and weakly on \textit{enrollment}. We interpret this as a dimension of socioeconomic disadvantage. For the purpose of the method, we ask, what would happen to the estimate of the PATE if the policy were deployed in a population more disadvantaged than the average in the experimental population? 

We can formalize this as a \textit{secular shift} in the positive direction of the first principal component. The direction $\theta$ is chosen so that the shift occurs along PC1. I vary its extent through simulation, and the norm of $\theta$ describes the size of the shift in terms of standard deviations along PC1. 

To summarize, our approach is as follows.

\begin{table}[H]\centering
\caption{Simulation design for the STAR re-analysis.}
\label{tab:star_design}
\begin{tabular}{@{}l p{0.78\textwidth}@{}}
\toprule
Step & Description \\
\midrule
1. Candidate set & Draw 40 schools at random from the 75 STAR schools. \\
\addlinespace
2. Selection & Each method selects $K$ schools from the candidate set. \\
\addlinespace
3. Deployment set & Shift the 35 held-out schools by $\theta^*$, a translation of $\|\theta^*\|$ standard deviations along the first principal component. The shifted schools are the deployment population $Q$. \\
\addlinespace
4. Estimate & Run the experiment in the selected schools, at their candidate-set covariates. Compare the estimate of the PATE from the selected sites to the PATE in $Q$. \\
\addlinespace
5. Report & Mean absolute bias of the PATE estimate, its mean squared error, and the mean $W_1$ distance from the selected sites to the deployment population\\
\bottomrule
\end{tabular}
\end{table}
\begin{figure}[H]\centering
\label{fig:star_design}
\begin{subfigure}{\textwidth}\centering
\resizebox{0.74\textwidth}{!}{%
\begin{tikzpicture}
  \draw[->, gray!65, line width=0.6pt] (-3.85,0) -- (8.47,0) node[right, font=\scriptsize, text=gray!75!black] {PC$_1$ -- disadvantage};
  \draw[->, gray!65, line width=0.6pt] (0,-3.27) -- (0,8.99) node[above, font=\scriptsize, text=gray!75!black] {PC$_2$};
  \fill[black!45, opacity=0.85] (-1.775,-1.297) circle (2.6pt);
  \fill[black!45, opacity=0.85] (-2.609,-0.847) circle (2.6pt);
  \fill[black!45, opacity=0.85] (-1.361,-0.617) circle (2.6pt);
  \fill[black!45, opacity=0.85] (-2.428,-0.075) circle (2.6pt);
  \fill[black!45, opacity=0.85] (-0.016,1.445) circle (2.6pt);
  \fill[black!45, opacity=0.85] (-0.110,-1.665) circle (2.6pt);
  \fill[black!45, opacity=0.85] (-2.611,-1.762) circle (2.6pt);
  \fill[black!45, opacity=0.85] (0.226,-2.423) circle (2.6pt);
  \fill[black!45, opacity=0.85] (-0.888,0.524) circle (2.6pt);
  \fill[black!45, opacity=0.85] (0.804,8.193) circle (2.6pt);
  \fill[black!45, opacity=0.85] (0.844,-0.353) circle (2.6pt);
  \fill[black!45, opacity=0.85] (-0.876,-0.496) circle (2.6pt);
  \fill[black!45, opacity=0.85] (-0.885,-0.425) circle (2.6pt);
  \fill[black!45, opacity=0.85] (-0.172,-1.319) circle (2.6pt);
  \fill[black!45, opacity=0.85] (-2.053,0.268) circle (2.6pt);
  \fill[black!45, opacity=0.85] (-0.917,-1.072) circle (2.6pt);
  \fill[black!45, opacity=0.85] (-0.823,3.012) circle (2.6pt);
  \fill[black!45, opacity=0.85] (-1.326,0.975) circle (2.6pt);
  \fill[black!45, opacity=0.85] (-1.350,-1.229) circle (2.6pt);
  \fill[black!45, opacity=0.85] (-2.007,-0.762) circle (2.6pt);
  \fill[black!45, opacity=0.85] (3.358,-0.302) circle (2.6pt);
  \fill[black!45, opacity=0.85] (-1.712,2.465) circle (2.6pt);
  \fill[black!45, opacity=0.85] (-1.379,-0.665) circle (2.6pt);
  \fill[black!45, opacity=0.85] (-1.997,0.109) circle (2.6pt);
  \fill[black!45, opacity=0.85] (-1.478,-0.332) circle (2.6pt);
  \fill[black!45, opacity=0.85] (-1.082,-1.531) circle (2.6pt);
  \fill[black!45, opacity=0.85] (-1.387,1.151) circle (2.6pt);
  \fill[black!45, opacity=0.85] (3.477,-2.072) circle (2.6pt);
  \fill[black!45, opacity=0.85] (4.092,2.033) circle (2.6pt);
  \fill[black!45, opacity=0.85] (3.565,-2.040) circle (2.6pt);
  \fill[black!45, opacity=0.85] (-1.364,0.496) circle (2.6pt);
  \fill[black!45, opacity=0.85] (-1.361,-1.920) circle (2.6pt);
  \fill[black!45, opacity=0.85] (-1.089,3.040) circle (2.6pt);
  \fill[black!45, opacity=0.85] (-2.933,1.144) circle (2.6pt);
  \fill[black!45, opacity=0.85] (-2.703,2.926) circle (2.6pt);
  \fill[blue!70!black, opacity=0.85] (1.092,-1.578) circle (2.8pt);
  \fill[blue!70!black, opacity=0.85] (-1.151,-0.247) circle (2.8pt);
  \fill[blue!70!black, opacity=0.85] (-0.795,0.526) circle (2.8pt);
  \fill[blue!70!black, opacity=0.85] (-1.872,-1.417) circle (2.8pt);
  \fill[blue!70!black, opacity=0.85] (-2.198,-0.887) circle (2.8pt);
  \fill[blue!70!black, opacity=0.85] (-0.431,-0.334) circle (2.8pt);
  \fill[blue!70!black, opacity=0.85] (-0.299,1.688) circle (2.8pt);
  \fill[blue!70!black, opacity=0.85] (-3.050,0.404) circle (2.8pt);
  \fill[blue!70!black, opacity=0.85] (-1.241,-0.941) circle (2.8pt);
  \fill[blue!70!black, opacity=0.85] (-0.000,-2.190) circle (2.8pt);
  \fill[blue!70!black, opacity=0.85] (-0.695,0.042) circle (2.8pt);
  \fill[blue!70!black, opacity=0.85] (3.802,-2.096) circle (2.8pt);
  \fill[blue!70!black, opacity=0.85] (-2.653,-0.990) circle (2.8pt);
  \fill[blue!70!black, opacity=0.85] (-2.032,-0.304) circle (2.8pt);
  \fill[blue!70!black, opacity=0.85] (-1.513,-1.904) circle (2.8pt);
  \fill[blue!70!black, opacity=0.85] (-1.438,2.354) circle (2.8pt);
  \fill[blue!70!black, opacity=0.85] (-0.271,0.587) circle (2.8pt);
  \fill[blue!70!black, opacity=0.85] (-0.964,1.899) circle (2.8pt);
  \fill[blue!70!black, opacity=0.85] (-0.439,1.333) circle (2.8pt);
  \fill[blue!70!black, opacity=0.85] (-2.302,-0.734) circle (2.8pt);
  \fill[blue!70!black, opacity=0.85] (-0.351,-1.342) circle (2.8pt);
  \fill[blue!70!black, opacity=0.85] (4.269,1.303) circle (2.8pt);
  \fill[blue!70!black, opacity=0.85] (4.414,1.500) circle (2.8pt);
  \fill[blue!70!black, opacity=0.85] (4.356,1.029) circle (2.8pt);
  \fill[blue!70!black, opacity=0.85] (1.720,1.149) circle (2.8pt);
  \fill[blue!70!black, opacity=0.85] (3.738,-1.892) circle (2.8pt);
  \fill[blue!70!black, opacity=0.85] (0.960,-0.203) circle (2.8pt);
  \fill[blue!70!black, opacity=0.85] (3.832,-1.371) circle (2.8pt);
  \fill[blue!70!black, opacity=0.85] (3.699,-2.573) circle (2.8pt);
  \fill[blue!70!black, opacity=0.85] (4.302,1.435) circle (2.8pt);
  \fill[blue!70!black, opacity=0.85] (3.786,0.619) circle (2.8pt);
  \fill[blue!70!black, opacity=0.85] (3.953,-2.084) circle (2.8pt);
  \fill[blue!70!black, opacity=0.85] (1.895,0.449) circle (2.8pt);
  \fill[blue!70!black, opacity=0.85] (0.763,-0.136) circle (2.8pt);
  \fill[blue!70!black, opacity=0.85] (3.987,2.251) circle (2.8pt);
  \fill[blue!70!black, opacity=0.85] (1.025,-0.888) circle (2.8pt);
  \fill[blue!70!black, opacity=0.85] (1.755,1.629) circle (2.8pt);
  \fill[blue!70!black, opacity=0.85] (-1.301,0.214) circle (2.8pt);
  \fill[blue!70!black, opacity=0.85] (-1.528,-0.818) circle (2.8pt);
  \fill[blue!70!black, opacity=0.85] (-2.495,-0.056) circle (2.8pt);
\end{tikzpicture}
}
\caption{Candidate set (blue, $n = 40$) and held-out set (grey, $n = 35$).}
\label{fig:star_design_a}
\end{subfigure}

\vspace{1.2em}
\begin{subfigure}{\textwidth}\centering
\resizebox{0.74\textwidth}{!}{%
\begin{tikzpicture}
  \draw[->, gray!65, line width=0.6pt] (-3.85,0) -- (8.47,0) node[right, font=\scriptsize, text=gray!75!black] {PC$_1$ -- disadvantage};
  \draw[->, gray!65, line width=0.6pt] (0,-3.27) -- (0,8.99) node[above, font=\scriptsize, text=gray!75!black] {PC$_2$};
  \fill[blue!70!black, opacity=0.35] (1.092,-1.578) circle (2.6pt);
  \fill[blue!70!black, opacity=0.35] (-1.151,-0.247) circle (2.6pt);
  \fill[blue!70!black, opacity=0.35] (-0.795,0.526) circle (2.6pt);
  \fill[blue!70!black, opacity=0.35] (-1.872,-1.417) circle (2.6pt);
  \fill[blue!70!black, opacity=0.35] (-2.198,-0.887) circle (2.6pt);
  \fill[blue!70!black, opacity=0.35] (-0.431,-0.334) circle (2.6pt);
  \fill[blue!70!black, opacity=0.35] (-0.299,1.688) circle (2.6pt);
  \fill[blue!70!black, opacity=0.35] (-3.050,0.404) circle (2.6pt);
  \fill[blue!70!black, opacity=0.35] (-1.241,-0.941) circle (2.6pt);
  \fill[blue!70!black, opacity=0.35] (-0.000,-2.190) circle (2.6pt);
  \fill[blue!70!black, opacity=0.35] (-0.695,0.042) circle (2.6pt);
  \fill[blue!70!black, opacity=0.35] (3.802,-2.096) circle (2.6pt);
  \fill[blue!70!black, opacity=0.35] (-2.653,-0.990) circle (2.6pt);
  \fill[blue!70!black, opacity=0.35] (-2.032,-0.304) circle (2.6pt);
  \fill[blue!70!black, opacity=0.35] (-1.513,-1.904) circle (2.6pt);
  \fill[blue!70!black, opacity=0.35] (-1.438,2.354) circle (2.6pt);
  \fill[blue!70!black, opacity=0.35] (-0.271,0.587) circle (2.6pt);
  \fill[blue!70!black, opacity=0.35] (-0.964,1.899) circle (2.6pt);
  \fill[blue!70!black, opacity=0.35] (-0.439,1.333) circle (2.6pt);
  \fill[blue!70!black, opacity=0.35] (-2.302,-0.734) circle (2.6pt);
  \fill[blue!70!black, opacity=0.35] (-0.351,-1.342) circle (2.6pt);
  \fill[blue!70!black, opacity=0.35] (4.269,1.303) circle (2.6pt);
  \fill[blue!70!black, opacity=0.35] (4.414,1.500) circle (2.6pt);
  \fill[blue!70!black, opacity=0.35] (4.356,1.029) circle (2.6pt);
  \fill[blue!70!black, opacity=0.35] (1.720,1.149) circle (2.6pt);
  \fill[blue!70!black, opacity=0.35] (3.738,-1.892) circle (2.6pt);
  \fill[blue!70!black, opacity=0.35] (0.960,-0.203) circle (2.6pt);
  \fill[blue!70!black, opacity=0.35] (3.832,-1.371) circle (2.6pt);
  \fill[blue!70!black, opacity=0.35] (3.699,-2.573) circle (2.6pt);
  \fill[blue!70!black, opacity=0.35] (4.302,1.435) circle (2.6pt);
  \fill[blue!70!black, opacity=0.35] (3.786,0.619) circle (2.6pt);
  \fill[blue!70!black, opacity=0.35] (3.953,-2.084) circle (2.6pt);
  \fill[blue!70!black, opacity=0.35] (1.895,0.449) circle (2.6pt);
  \fill[blue!70!black, opacity=0.35] (0.763,-0.136) circle (2.6pt);
  \fill[blue!70!black, opacity=0.35] (3.987,2.251) circle (2.6pt);
  \fill[blue!70!black, opacity=0.35] (1.025,-0.888) circle (2.6pt);
  \fill[blue!70!black, opacity=0.35] (1.755,1.629) circle (2.6pt);
  \fill[blue!70!black, opacity=0.35] (-1.301,0.214) circle (2.6pt);
  \fill[blue!70!black, opacity=0.35] (-1.528,-0.818) circle (2.6pt);
  \fill[blue!70!black, opacity=0.35] (-2.495,-0.056) circle (2.6pt);
  \fill[black!35, opacity=0.55] (-1.775,-1.297) circle (2.4pt);
  \fill[black!35, opacity=0.55] (-2.609,-0.847) circle (2.4pt);
  \fill[black!35, opacity=0.55] (-1.361,-0.617) circle (2.4pt);
  \fill[black!35, opacity=0.55] (-2.428,-0.075) circle (2.4pt);
  \fill[black!35, opacity=0.55] (-0.016,1.445) circle (2.4pt);
  \fill[black!35, opacity=0.55] (-0.110,-1.665) circle (2.4pt);
  \fill[black!35, opacity=0.55] (-2.611,-1.762) circle (2.4pt);
  \fill[black!35, opacity=0.55] (0.226,-2.423) circle (2.4pt);
  \fill[black!35, opacity=0.55] (-0.888,0.524) circle (2.4pt);
  \fill[black!35, opacity=0.55] (0.804,8.193) circle (2.4pt);
  \fill[black!35, opacity=0.55] (0.844,-0.353) circle (2.4pt);
  \fill[black!35, opacity=0.55] (-0.876,-0.496) circle (2.4pt);
  \fill[black!35, opacity=0.55] (-0.885,-0.425) circle (2.4pt);
  \fill[black!35, opacity=0.55] (-0.172,-1.319) circle (2.4pt);
  \fill[black!35, opacity=0.55] (-2.053,0.268) circle (2.4pt);
  \fill[black!35, opacity=0.55] (-0.917,-1.072) circle (2.4pt);
  \fill[black!35, opacity=0.55] (-0.823,3.012) circle (2.4pt);
  \fill[black!35, opacity=0.55] (-1.326,0.975) circle (2.4pt);
  \fill[black!35, opacity=0.55] (-1.350,-1.229) circle (2.4pt);
  \fill[black!35, opacity=0.55] (-2.007,-0.762) circle (2.4pt);
  \fill[black!35, opacity=0.55] (3.358,-0.302) circle (2.4pt);
  \fill[black!35, opacity=0.55] (-1.712,2.465) circle (2.4pt);
  \fill[black!35, opacity=0.55] (-1.379,-0.665) circle (2.4pt);
  \fill[black!35, opacity=0.55] (-1.997,0.109) circle (2.4pt);
  \fill[black!35, opacity=0.55] (-1.478,-0.332) circle (2.4pt);
  \fill[black!35, opacity=0.55] (-1.082,-1.531) circle (2.4pt);
  \fill[black!35, opacity=0.55] (-1.387,1.151) circle (2.4pt);
  \fill[black!35, opacity=0.55] (3.477,-2.072) circle (2.4pt);
  \fill[black!35, opacity=0.55] (4.092,2.033) circle (2.4pt);
  \fill[black!35, opacity=0.55] (3.565,-2.040) circle (2.4pt);
  \fill[black!35, opacity=0.55] (-1.364,0.496) circle (2.4pt);
  \fill[black!35, opacity=0.55] (-1.361,-1.920) circle (2.4pt);
  \fill[black!35, opacity=0.55] (-1.089,3.040) circle (2.4pt);
  \fill[black!35, opacity=0.55] (-2.933,1.144) circle (2.4pt);
  \fill[black!35, opacity=0.55] (-2.703,2.926) circle (2.4pt);
  \fill[green!45!black, opacity=0.85] (1.601,-1.297) circle (2.8pt);
  \fill[green!45!black, opacity=0.85] (0.767,-0.847) circle (2.8pt);
  \fill[green!45!black, opacity=0.85] (2.016,-0.617) circle (2.8pt);
  \fill[green!45!black, opacity=0.85] (0.948,-0.075) circle (2.8pt);
  \fill[green!45!black, opacity=0.85] (3.360,1.445) circle (2.8pt);
  \fill[green!45!black, opacity=0.85] (3.266,-1.665) circle (2.8pt);
  \fill[green!45!black, opacity=0.85] (0.765,-1.762) circle (2.8pt);
  \fill[green!45!black, opacity=0.85] (3.602,-2.423) circle (2.8pt);
  \fill[green!45!black, opacity=0.85] (2.488,0.524) circle (2.8pt);
  \fill[green!45!black, opacity=0.85] (4.180,8.193) circle (2.8pt);
  \fill[green!45!black, opacity=0.85] (4.220,-0.353) circle (2.8pt);
  \fill[green!45!black, opacity=0.85] (2.500,-0.496) circle (2.8pt);
  \fill[green!45!black, opacity=0.85] (2.491,-0.425) circle (2.8pt);
  \fill[green!45!black, opacity=0.85] (3.204,-1.319) circle (2.8pt);
  \fill[green!45!black, opacity=0.85] (1.323,0.268) circle (2.8pt);
  \fill[green!45!black, opacity=0.85] (2.459,-1.072) circle (2.8pt);
  \fill[green!45!black, opacity=0.85] (2.553,3.012) circle (2.8pt);
  \fill[green!45!black, opacity=0.85] (2.050,0.975) circle (2.8pt);
  \fill[green!45!black, opacity=0.85] (2.026,-1.229) circle (2.8pt);
  \fill[green!45!black, opacity=0.85] (1.369,-0.762) circle (2.8pt);
  \fill[green!45!black, opacity=0.85] (6.734,-0.302) circle (2.8pt);
  \fill[green!45!black, opacity=0.85] (1.664,2.465) circle (2.8pt);
  \fill[green!45!black, opacity=0.85] (1.997,-0.665) circle (2.8pt);
  \fill[green!45!black, opacity=0.85] (1.379,0.109) circle (2.8pt);
  \fill[green!45!black, opacity=0.85] (1.898,-0.332) circle (2.8pt);
  \fill[green!45!black, opacity=0.85] (2.294,-1.531) circle (2.8pt);
  \fill[green!45!black, opacity=0.85] (1.989,1.151) circle (2.8pt);
  \fill[green!45!black, opacity=0.85] (6.853,-2.072) circle (2.8pt);
  \fill[green!45!black, opacity=0.85] (7.468,2.033) circle (2.8pt);
  \fill[green!45!black, opacity=0.85] (6.941,-2.040) circle (2.8pt);
  \fill[green!45!black, opacity=0.85] (2.012,0.496) circle (2.8pt);
  \fill[green!45!black, opacity=0.85] (2.015,-1.920) circle (2.8pt);
  \fill[green!45!black, opacity=0.85] (2.287,3.040) circle (2.8pt);
  \fill[green!45!black, opacity=0.85] (0.443,1.144) circle (2.8pt);
  \fill[green!45!black, opacity=0.85] (0.673,2.926) circle (2.8pt);
  \draw[->, green!35!black, line width=1.6pt] (-0.695,0.131) -- (2.681,0.131);
  \node[font=\small, green!30!black] at (0.993,0.681) {$\theta$};
  \draw[black!55, line width=1.5pt] (-0.695,0.131) ++(-5pt,-5pt) -- ++(10pt,10pt);
  \draw[black!55, line width=1.5pt] (-0.695,0.131) ++(-5pt,5pt) -- ++(10pt,-10pt);
  \draw[green!35!black, line width=1.5pt] (2.681,0.131) ++(-5pt,-5pt) -- ++(10pt,10pt);
  \draw[green!35!black, line width=1.5pt] (2.681,0.131) ++(-5pt,5pt) -- ++(10pt,-10pt);
\end{tikzpicture}
}
\caption{Translation shift applied to the held-out schools. Crosses mark the shift from $ATE(P)$ to $ATE (Q_\theta)$.}
\label{fig:star_design_b}
\end{subfigure}
\caption{Simulation design. The sites are randomly partitioned into candidate sites and held-out sites (top). Held-out sites are then shifted adversarially in the direction of $\theta$ (bottom).}
\end{figure}
\begin{figure}[H]\centering
\label{fig:star_selection}
\resizebox{0.84\textwidth}{!}{%
\begin{tikzpicture}
  \draw[->, gray!65, line width=0.6pt] (-3.85,0) -- (8.47,0) node[right, font=\scriptsize, text=gray!75!black] {PC$_1$ -- disadvantage};
  \draw[->, gray!65, line width=0.6pt] (0,-3.27) -- (0,8.99) node[above, font=\scriptsize, text=gray!75!black] {PC$_2$};
  \fill[black!25, opacity=0.50] (-1.775,-1.297) circle (2.2pt);
  \fill[black!25, opacity=0.50] (-2.609,-0.847) circle (2.2pt);
  \fill[black!25, opacity=0.50] (-1.361,-0.617) circle (2.2pt);
  \fill[black!25, opacity=0.50] (-2.428,-0.075) circle (2.2pt);
  \fill[black!25, opacity=0.50] (-0.016,1.445) circle (2.2pt);
  \fill[black!25, opacity=0.50] (-0.110,-1.665) circle (2.2pt);
  \fill[black!25, opacity=0.50] (-2.611,-1.762) circle (2.2pt);
  \fill[black!25, opacity=0.50] (0.226,-2.423) circle (2.2pt);
  \fill[black!25, opacity=0.50] (-0.888,0.524) circle (2.2pt);
  \fill[black!25, opacity=0.50] (0.804,8.193) circle (2.2pt);
  \fill[black!25, opacity=0.50] (0.844,-0.353) circle (2.2pt);
  \fill[black!25, opacity=0.50] (-0.876,-0.496) circle (2.2pt);
  \fill[black!25, opacity=0.50] (-0.885,-0.425) circle (2.2pt);
  \fill[black!25, opacity=0.50] (-0.172,-1.319) circle (2.2pt);
  \fill[black!25, opacity=0.50] (-2.053,0.268) circle (2.2pt);
  \fill[black!25, opacity=0.50] (-0.917,-1.072) circle (2.2pt);
  \fill[black!25, opacity=0.50] (-0.823,3.012) circle (2.2pt);
  \fill[black!25, opacity=0.50] (-1.326,0.975) circle (2.2pt);
  \fill[black!25, opacity=0.50] (-1.350,-1.229) circle (2.2pt);
  \fill[black!25, opacity=0.50] (-2.007,-0.762) circle (2.2pt);
  \fill[black!25, opacity=0.50] (3.358,-0.302) circle (2.2pt);
  \fill[black!25, opacity=0.50] (-1.712,2.465) circle (2.2pt);
  \fill[black!25, opacity=0.50] (-1.379,-0.665) circle (2.2pt);
  \fill[black!25, opacity=0.50] (-1.997,0.109) circle (2.2pt);
  \fill[black!25, opacity=0.50] (-1.478,-0.332) circle (2.2pt);
  \fill[black!25, opacity=0.50] (-1.082,-1.531) circle (2.2pt);
  \fill[black!25, opacity=0.50] (-1.387,1.151) circle (2.2pt);
  \fill[black!25, opacity=0.50] (3.477,-2.072) circle (2.2pt);
  \fill[black!25, opacity=0.50] (4.092,2.033) circle (2.2pt);
  \fill[black!25, opacity=0.50] (3.565,-2.040) circle (2.2pt);
  \fill[black!25, opacity=0.50] (-1.364,0.496) circle (2.2pt);
  \fill[black!25, opacity=0.50] (-1.361,-1.920) circle (2.2pt);
  \fill[black!25, opacity=0.50] (-1.089,3.040) circle (2.2pt);
  \fill[black!25, opacity=0.50] (-2.933,1.144) circle (2.2pt);
  \fill[black!25, opacity=0.50] (-2.703,2.926) circle (2.2pt);
  \fill[green!45!black, opacity=0.35] (1.601,-1.297) circle (2.4pt);
  \fill[green!45!black, opacity=0.35] (0.767,-0.847) circle (2.4pt);
  \fill[green!45!black, opacity=0.35] (2.016,-0.617) circle (2.4pt);
  \fill[green!45!black, opacity=0.35] (0.948,-0.075) circle (2.4pt);
  \fill[green!45!black, opacity=0.35] (3.360,1.445) circle (2.4pt);
  \fill[green!45!black, opacity=0.35] (3.266,-1.665) circle (2.4pt);
  \fill[green!45!black, opacity=0.35] (0.765,-1.762) circle (2.4pt);
  \fill[green!45!black, opacity=0.35] (3.602,-2.423) circle (2.4pt);
  \fill[green!45!black, opacity=0.35] (2.488,0.524) circle (2.4pt);
  \fill[green!45!black, opacity=0.35] (4.180,8.193) circle (2.4pt);
  \fill[green!45!black, opacity=0.35] (4.220,-0.353) circle (2.4pt);
  \fill[green!45!black, opacity=0.35] (2.500,-0.496) circle (2.4pt);
  \fill[green!45!black, opacity=0.35] (2.491,-0.425) circle (2.4pt);
  \fill[green!45!black, opacity=0.35] (3.204,-1.319) circle (2.4pt);
  \fill[green!45!black, opacity=0.35] (1.323,0.268) circle (2.4pt);
  \fill[green!45!black, opacity=0.35] (2.459,-1.072) circle (2.4pt);
  \fill[green!45!black, opacity=0.35] (2.553,3.012) circle (2.4pt);
  \fill[green!45!black, opacity=0.35] (2.050,0.975) circle (2.4pt);
  \fill[green!45!black, opacity=0.35] (2.026,-1.229) circle (2.4pt);
  \fill[green!45!black, opacity=0.35] (1.369,-0.762) circle (2.4pt);
  \fill[green!45!black, opacity=0.35] (6.734,-0.302) circle (2.4pt);
  \fill[green!45!black, opacity=0.35] (1.664,2.465) circle (2.4pt);
  \fill[green!45!black, opacity=0.35] (1.997,-0.665) circle (2.4pt);
  \fill[green!45!black, opacity=0.35] (1.379,0.109) circle (2.4pt);
  \fill[green!45!black, opacity=0.35] (1.898,-0.332) circle (2.4pt);
  \fill[green!45!black, opacity=0.35] (2.294,-1.531) circle (2.4pt);
  \fill[green!45!black, opacity=0.35] (1.989,1.151) circle (2.4pt);
  \fill[green!45!black, opacity=0.35] (6.853,-2.072) circle (2.4pt);
  \fill[green!45!black, opacity=0.35] (7.468,2.033) circle (2.4pt);
  \fill[green!45!black, opacity=0.35] (6.941,-2.040) circle (2.4pt);
  \fill[green!45!black, opacity=0.35] (2.012,0.496) circle (2.4pt);
  \fill[green!45!black, opacity=0.35] (2.015,-1.920) circle (2.4pt);
  \fill[green!45!black, opacity=0.35] (2.287,3.040) circle (2.4pt);
  \fill[green!45!black, opacity=0.35] (0.443,1.144) circle (2.4pt);
  \fill[green!45!black, opacity=0.35] (0.673,2.926) circle (2.4pt);
  \fill[blue!70!black, opacity=0.55] (1.092,-1.578) circle (2.6pt);
  \fill[blue!70!black, opacity=0.55] (-1.151,-0.247) circle (2.6pt);
  \fill[blue!70!black, opacity=0.55] (-0.795,0.526) circle (2.6pt);
  \fill[blue!70!black, opacity=0.55] (-1.872,-1.417) circle (2.6pt);
  \fill[blue!70!black, opacity=0.55] (-2.198,-0.887) circle (2.6pt);
  \fill[blue!70!black, opacity=0.55] (-0.431,-0.334) circle (2.6pt);
  \fill[blue!70!black, opacity=0.55] (-0.299,1.688) circle (2.6pt);
  \fill[blue!70!black, opacity=0.55] (-3.050,0.404) circle (2.6pt);
  \fill[blue!70!black, opacity=0.55] (-1.241,-0.941) circle (2.6pt);
  \fill[blue!70!black, opacity=0.55] (-0.000,-2.190) circle (2.6pt);
  \fill[blue!70!black, opacity=0.55] (-0.695,0.042) circle (2.6pt);
  \fill[blue!70!black, opacity=0.55] (3.802,-2.096) circle (2.6pt);
  \fill[blue!70!black, opacity=0.55] (-2.653,-0.990) circle (2.6pt);
  \fill[blue!70!black, opacity=0.55] (-2.032,-0.304) circle (2.6pt);
  \fill[blue!70!black, opacity=0.55] (-1.513,-1.904) circle (2.6pt);
  \fill[blue!70!black, opacity=0.55] (-1.438,2.354) circle (2.6pt);
  \fill[blue!70!black, opacity=0.55] (-0.271,0.587) circle (2.6pt);
  \fill[blue!70!black, opacity=0.55] (-0.964,1.899) circle (2.6pt);
  \fill[blue!70!black, opacity=0.55] (-0.439,1.333) circle (2.6pt);
  \fill[blue!70!black, opacity=0.55] (-2.302,-0.734) circle (2.6pt);
  \fill[blue!70!black, opacity=0.55] (-0.351,-1.342) circle (2.6pt);
  \fill[blue!70!black, opacity=0.55] (4.269,1.303) circle (2.6pt);
  \fill[blue!70!black, opacity=0.55] (4.414,1.500) circle (2.6pt);
  \fill[blue!70!black, opacity=0.55] (4.356,1.029) circle (2.6pt);
  \fill[blue!70!black, opacity=0.55] (1.720,1.149) circle (2.6pt);
  \fill[blue!70!black, opacity=0.55] (3.738,-1.892) circle (2.6pt);
  \fill[blue!70!black, opacity=0.55] (0.960,-0.203) circle (2.6pt);
  \fill[blue!70!black, opacity=0.55] (3.832,-1.371) circle (2.6pt);
  \fill[blue!70!black, opacity=0.55] (3.699,-2.573) circle (2.6pt);
  \fill[blue!70!black, opacity=0.55] (4.302,1.435) circle (2.6pt);
  \fill[blue!70!black, opacity=0.55] (3.786,0.619) circle (2.6pt);
  \fill[blue!70!black, opacity=0.55] (3.953,-2.084) circle (2.6pt);
  \fill[blue!70!black, opacity=0.55] (1.895,0.449) circle (2.6pt);
  \fill[blue!70!black, opacity=0.55] (0.763,-0.136) circle (2.6pt);
  \fill[blue!70!black, opacity=0.55] (3.987,2.251) circle (2.6pt);
  \fill[blue!70!black, opacity=0.55] (1.025,-0.888) circle (2.6pt);
  \fill[blue!70!black, opacity=0.55] (1.755,1.629) circle (2.6pt);
  \fill[blue!70!black, opacity=0.55] (-1.301,0.214) circle (2.6pt);
  \fill[blue!70!black, opacity=0.55] (-1.528,-0.818) circle (2.6pt);
  \fill[blue!70!black, opacity=0.55] (-2.495,-0.056) circle (2.6pt);
  \fill[red!75!black, opacity=0.95] (4.356,1.029) circle (4.4pt);
  \fill[red!75!black, opacity=0.95] (0.960,-0.203) circle (4.4pt);
  \fill[red!75!black, opacity=0.95] (3.953,-2.084) circle (4.4pt);
  \fill[red!75!black, opacity=0.95] (1.025,-0.888) circle (4.4pt);
  \fill[red!75!black, opacity=0.95] (1.755,1.629) circle (4.4pt);
  \draw[green!35!black, line width=1.5pt] (2.681,0.131) ++(-5pt,-5pt) -- ++(10pt,10pt);
  \draw[green!35!black, line width=1.5pt] (2.681,0.131) ++(-5pt,5pt) -- ++(10pt,-10pt);
   \draw[red!75!black, line width=1.5pt] (2.410,-0.103) ++(-6pt,-6pt) -- ++(12pt,12pt);
  \draw[red!75!black, line width=1.5pt] (2.410,-0.103) ++(-6pt,6pt) -- ++(12pt,-12pt);
\end{tikzpicture}
}
\label{fig:star_selection_a}
\caption{Sites selected by the Wasserstein DRO method under distribution shift. Green cross indicates $ATE(Q_\theta)$, red cross indicates $ATE(S^*_{\textrm{DRO}})$}
\end{figure}

\subsubsection*{Results}

In the above simulation, the Wasserstein-DRO method performs favourably under induced distribution shift, whenever distribution shift is greater than zero. It does so for the quantity it is explicitly intended to target, the $W_1$ distance, but also for the MSE and bias as well, which it upper bounds. 

The method of \citet{olea2024externallyvalidselectionexperimental} outperforms WDRO in the absence of distribution shift. This is in part because WDRO pays `the price of robustness' \citep{Bertsimas_sim_2004_priceofrobustness} -- a method optimal for the robust counterpart of an optimization problem is generally not optimal for the nominal version of that optimization problem. 

The method has a structural advantage in this simulation, in that it `knows' the correct value of $\theta$. 

Further simulation studies are needed to assess the method i) when $\theta$ is misspecified ii) under different levels of covariate informativeness and ii) under different models of distribution shift.

\begin{figure}[H]\centering
\label{fig:drift}

\vspace{0.4em}
\begin{subfigure}{0.62\textwidth}\centering
\includegraphics[width=\linewidth]{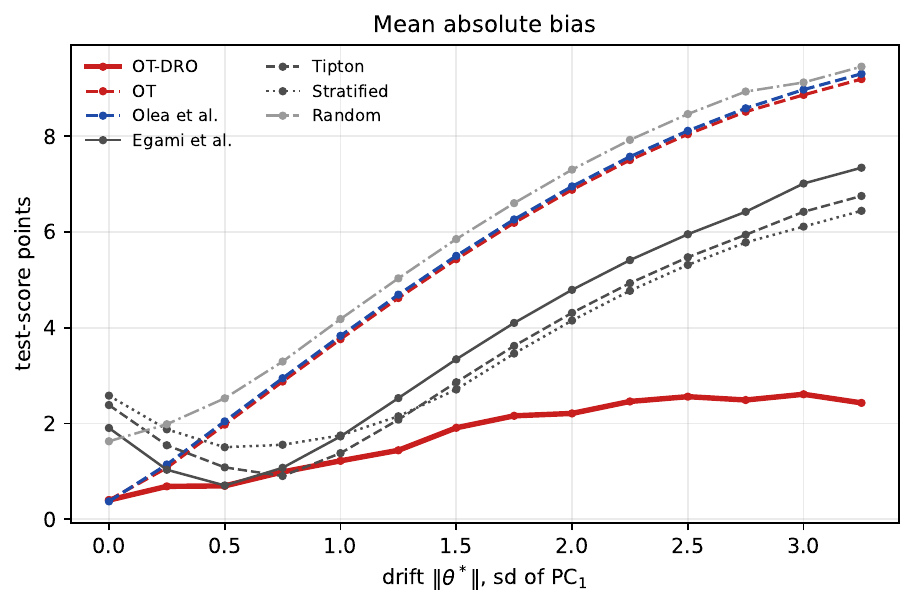}
\caption{Mean absolute bias of the deployment estimate.}
\label{fig:drift_bias}
\end{subfigure}

\vspace{0.8em}
\begin{subfigure}{0.62\textwidth}\centering
\includegraphics[width=\linewidth]{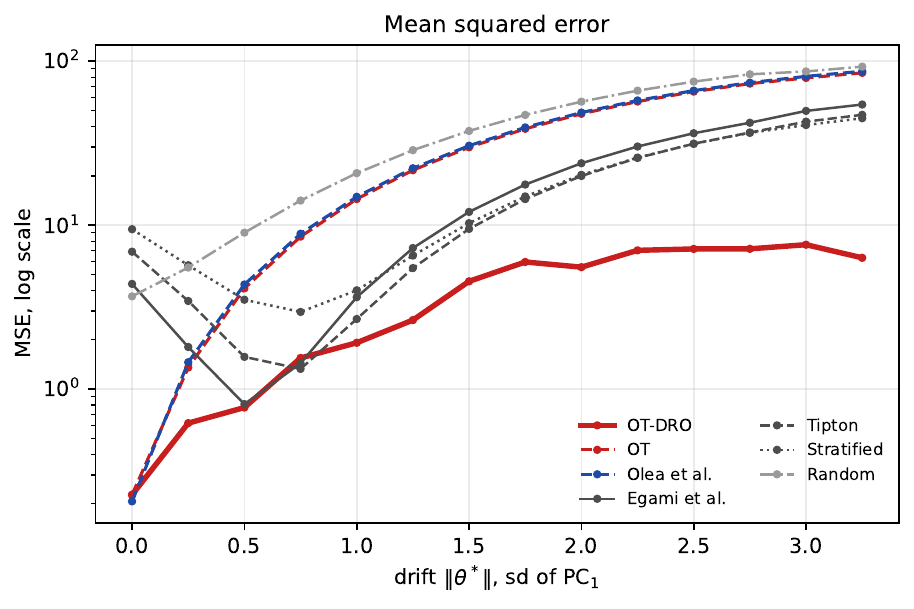}
\caption{Mean squared error of the deployment estimate, log scale.}
\label{fig:drift_mse}
\end{subfigure}

\vspace{0.8em}
\begin{subfigure}{0.62\textwidth}\centering
\includegraphics[width=\linewidth]{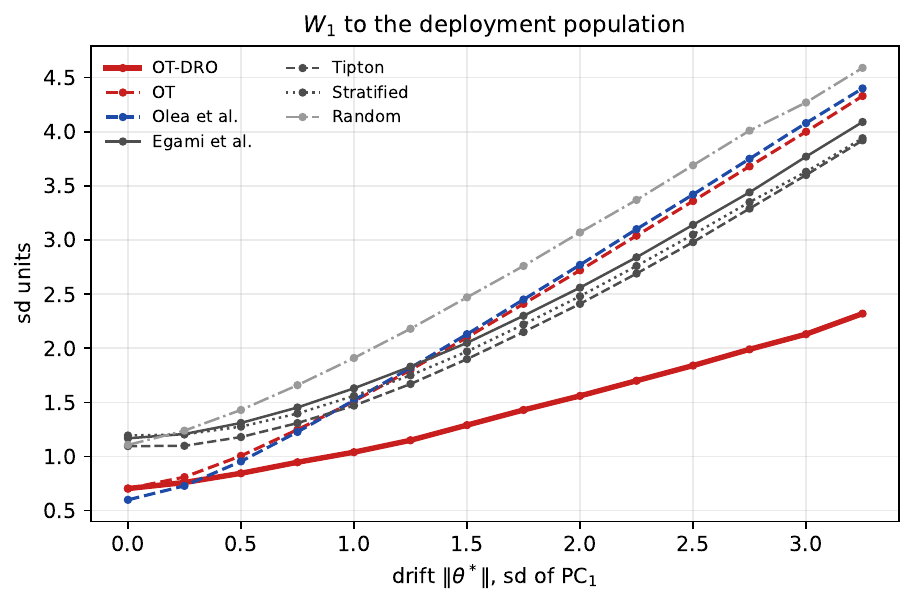}
\caption{Mean $W_1$ distance from the selected sites to the deployment population.}
\label{fig:drift_w1}
\end{subfigure}
\end{figure}

\begin{table}[H]\centering
\caption{Site selection under covariate drift}
\label{tab:drift}
\vspace{0.6em}
\begin{subtable}{\textwidth}\centering
\caption{Mean absolute bias of the deployment estimate, test-score points.}
\label{tab:drift_bias}
\begin{tabular}{lrrrr}
\toprule
method & $\|\theta^*\| = 0$ & $1$ & $2$ & $3$ \\
\midrule
W-DRO & 0.40 & \textbf{1.22} & \textbf{2.21} & \textbf{2.61} \\
Olea et al. & \textbf{0.37} & 3.83 & 6.95 & 8.97 \\
Egami et al. & 1.91 & 1.73 & 4.79 & 7.01 \\
Tipton & 2.38 & 1.38 & 4.31 & 6.42 \\
Stratified & 2.58 & 1.75 & 4.15 & 6.11 \\
Random & 1.63 & 4.18 & 7.30 & 9.12 \\
\bottomrule
\end{tabular}
\end{subtable}

\vspace{1.4em}
\begin{subtable}{\textwidth}\centering
\caption{Mean squared error of the deployment estimate.}
\label{tab:drift_mse}
\begin{tabular}{lrrrr}
\toprule
method & $\|\theta^*\| = 0$ & $1$ & $2$ & $3$ \\
\midrule
W-DRO & \textbf{0.2} & \textbf{1.9} & \textbf{5.5} & \textbf{7.6} \\
Olea et al. & \textbf{0.2} & 14.9 & 48.7 & 80.9 \\
Egami et al. & 4.4 & 3.6 & 23.9 & 49.8 \\
Tipton & 6.9 & 2.7 & 20.0 & 42.8 \\
Stratified & 9.5 & 4.0 & 20.3 & 40.9 \\
Random & 3.7 & 20.8 & 56.7 & 86.7 \\
\bottomrule
\end{tabular}
\end{subtable}

\vspace{1.4em}
\begin{subtable}{\textwidth}\centering
\caption{Mean $W_1$ distance from the selected sites to the deployment population}
\label{tab:drift_w1}
\begin{tabular}{lrrrr}
\toprule
method & $\|\theta^*\| = 0$ & $1$ & $2$ & $3$ \\
\midrule
W-DRO & 0.70 & \textbf{1.04} & \textbf{1.56} & \textbf{2.13} \\
Olea et al. & \textbf{0.60} & 1.52 & 2.77 & 4.08 \\
Egami et al. & 1.17 & 1.63 & 2.56 & 3.77 \\
Tipton & 1.10 & 1.47 & 2.41 & 3.60 \\
Stratified & 1.19 & 1.56 & 2.48 & 3.63 \\
Random & 1.11 & 1.91 & 3.07 & 4.27 \\
\bottomrule
\end{tabular}
\end{subtable}
\end{table}

\subsection{Cr\'epon et al.: overweighting of high market potential villages}\label{sec:crepon}

\subsubsection*{Study Characteristics}

\begin{landscape}
\begin{figure}[H]\centering
\includegraphics[width=0.95\linewidth, height=0.82\textheight, keepaspectratio]{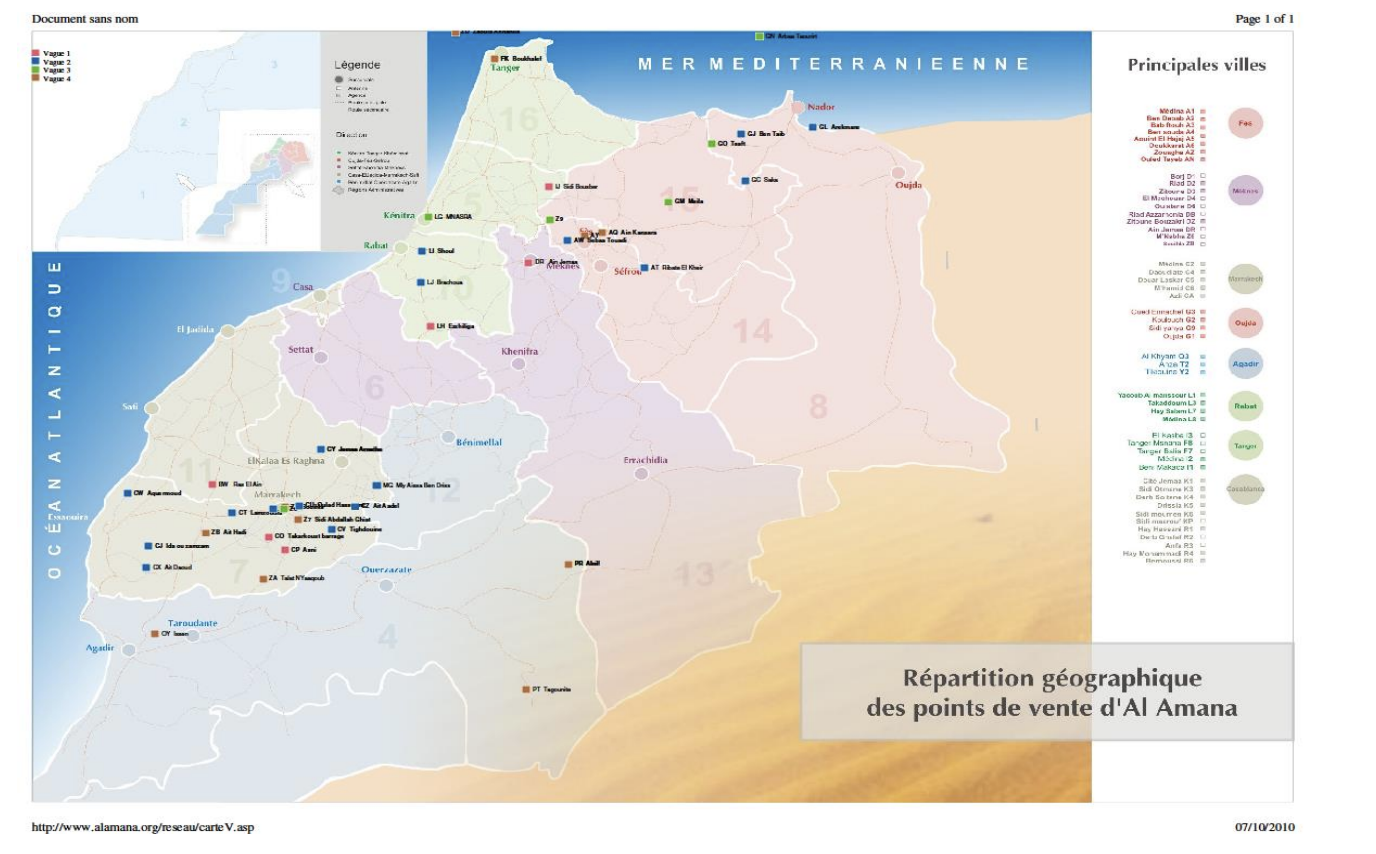}
\caption{Branch intervention areas in the Cr\'epon et al.\ study.}
\label{fig:maroc}
\end{figure}
\end{landscape}

\citet{Crepon_2015} studied the effect of an expansion in the availability of micocredit in rural areas in Morocco, when microfinance institution Al Amana expanded into a number of villages in 2006, on local farm profits and poverty reduction. The study population consisted of 162 villages in 81 matched pairs, where one village in each pair was randomly assigned to receive a new bank branch, which issued loans of a value between 1000 and 15000 Moroccan Dirhams (\$105 to \$1500). We take the matched pair as the experimental site. Each branch's service area was fixed by a map, shown in \Cref{fig:maroc}. Loan take-up was 13 \% in treated villages, and 0\% in untreated villages. 

The loans were found to increase self-employed profits by 2006 MAD against a control mean of 9056 MAD, an increase of around 22\%. Profit effects were heterogeneous across villages.  

In this exercise, we ask, what would the results of this study have been if the inclusion mechanism overweighted the market potential of the villages in the study? This is intended to recreate the problems described by \citet{Allcott_2015}, in which a site selection yields an overly optimistic estimate of the effect of a policy. 

\begin{figure}[htbp]\centering
\caption{Heterogeneous profits across villages.}
\label{fig:crepon_profit_hist}
\resizebox{0.9\textwidth}{!}{%
\begin{tikzpicture}[x=1cm, y=1cm]
  \fill[blue!70!black, opacity=0.55] (0.000,0.000) rectangle (0.650,0.333);
  \fill[blue!70!black, opacity=0.55] (1.950,0.000) rectangle (2.600,0.333);
  \fill[blue!70!black, opacity=0.55] (3.250,0.000) rectangle (3.900,0.333);
  \fill[blue!70!black, opacity=0.55] (3.900,0.000) rectangle (4.550,2.000);
  \fill[blue!70!black, opacity=0.55] (4.550,0.000) rectangle (5.200,7.000);
  \fill[blue!70!black, opacity=0.55] (5.200,0.000) rectangle (5.850,6.667);
  \fill[blue!70!black, opacity=0.55] (5.850,0.000) rectangle (6.500,3.667);
  \fill[blue!70!black, opacity=0.55] (6.500,0.000) rectangle (7.150,2.333);
  \fill[blue!70!black, opacity=0.55] (7.150,0.000) rectangle (7.800,2.000);
  \fill[blue!70!black, opacity=0.55] (7.800,0.000) rectangle (8.450,1.667);
  \fill[blue!70!black, opacity=0.55] (8.450,0.000) rectangle (9.100,0.667);
  \fill[red!75!black, opacity=0.55] (1.950,0.000) rectangle (2.600,0.333);
  \fill[red!75!black, opacity=0.55] (2.600,0.000) rectangle (3.250,0.333);
  \fill[red!75!black, opacity=0.55] (3.250,0.000) rectangle (3.900,1.000);
  \fill[red!75!black, opacity=0.55] (3.900,0.000) rectangle (4.550,3.333);
  \fill[red!75!black, opacity=0.55] (4.550,0.000) rectangle (5.200,6.333);
  \fill[red!75!black, opacity=0.55] (5.200,0.000) rectangle (5.850,3.667);
  \fill[red!75!black, opacity=0.55] (5.850,0.000) rectangle (6.500,5.667);
  \fill[red!75!black, opacity=0.55] (6.500,0.000) rectangle (7.150,1.667);
  \fill[red!75!black, opacity=0.55] (7.150,0.000) rectangle (7.800,1.333);
  \fill[red!75!black, opacity=0.55] (7.800,0.000) rectangle (8.450,0.667);
  \fill[red!75!black, opacity=0.55] (8.450,0.000) rectangle (9.100,0.667);
  \fill[red!75!black, opacity=0.55] (9.100,0.000) rectangle (9.750,0.333);
  \fill[red!75!black, opacity=0.55] (9.750,0.000) rectangle (10.400,0.333);
  \fill[red!75!black, opacity=0.55] (11.050,0.000) rectangle (11.700,0.333);
  \fill[red!75!black, opacity=0.55] (11.700,0.000) rectangle (12.350,0.667);
  \fill[red!75!black, opacity=0.55] (12.350,0.000) rectangle (13.000,0.333);
  \draw[blue!70!black, dashed, line width=1.2pt] (5.739,0) -- (5.739,7.000);
  \draw[red!75!black, dashed, line width=1.2pt] (5.969,0) -- (5.969,7.000);
  \draw[black!70, line width=0.7pt] (0,0) -- (13.000,0);
  \draw[black!70, line width=0.7pt] (0,0) -- (0,7.000);
  \draw[black!70] (1.950,0) -- (1.950,-0.12);
  \node[below, font=\scriptsize] at (1.950,-0.12) {-20};
  \draw[black!70] (4.550,0) -- (4.550,-0.12);
  \node[below, font=\scriptsize] at (4.550,-0.12) {0};
  \draw[black!70] (7.150,0) -- (7.150,-0.12);
  \node[below, font=\scriptsize] at (7.150,-0.12) {20};
  \draw[black!70] (9.750,0) -- (9.750,-0.12);
  \node[below, font=\scriptsize] at (9.750,-0.12) {40};
  \draw[black!70] (12.350,0) -- (12.350,-0.12);
  \node[below, font=\scriptsize] at (12.350,-0.12) {60};
  \draw[black!70] (0,0.000) -- (-0.12,0.000);
  \node[left, font=\scriptsize] at (-0.12,0.000) {0};
  \draw[black!70] (0,1.667) -- (-0.12,1.667);
  \node[left, font=\scriptsize] at (-0.12,1.667) {5};
  \draw[black!70] (0,3.333) -- (-0.12,3.333);
  \node[left, font=\scriptsize] at (-0.12,3.333) {10};
  \draw[black!70] (0,5.000) -- (-0.12,5.000);
  \node[left, font=\scriptsize] at (-0.12,5.000) {15};
  \draw[black!70] (0,6.667) -- (-0.12,6.667);
  \node[left, font=\scriptsize] at (-0.12,6.667) {20};
  \node[below, font=\small] at (6.500,-0.85) {village mean profit, thousands of MAD per year};
  \node[rotate=90, above, font=\small] at (-1.05,3.500) {villages};
  \fill[blue!70!black, opacity=0.55] (8.060,6.650) rectangle (8.510,6.970);
  \node[right, font=\scriptsize] at (8.610,6.810) {control, mean 9.1};
  \fill[red!75!black, opacity=0.55] (8.060,6.050) rectangle (8.510,6.370);
  \node[right, font=\scriptsize] at (8.610,6.210) {treated, mean 10.9};
\end{tikzpicture}
}
\end{figure}

\subsubsection*{Simulation Design}

To build an adversarial situation for this study, we project site-level covariates onto principal components, and construct a reweighting shift that upweights villages with lower market potential. Lower market potential includes the covariates: baseline business, agricultural and livestock activity, business assets, asset index, prior borrowing, and poverty (negatively signed). 

For this exercise we use the \textit{tilt} model of drift, which is a reweighting shift in the negative direction of market potential.  \Cref{fig:crepon_level_hist} shows that the applied transformation tilts the distribution of site-level profits to the left.

\begin{figure}[htbp]\centering
\caption{Tilted profits.}
\label{fig:crepon_level_hist}
\resizebox{0.88\textwidth}{!}{%
\begin{tikzpicture}[x=1cm, y=1cm]
  \fill[blue!70!black, opacity=0.45] (0.000,0) rectangle (0.929,0.180);
  \fill[blue!70!black, opacity=0.45] (2.786,0) rectangle (3.714,0.180);
  \fill[blue!70!black, opacity=0.45] (4.643,0) rectangle (5.571,0.180);
  \fill[blue!70!black, opacity=0.45] (5.571,0) rectangle (6.500,0.901);
  \fill[blue!70!black, opacity=0.45] (6.500,0) rectangle (7.429,3.424);
  \fill[blue!70!black, opacity=0.45] (7.429,0) rectangle (8.357,3.063);
  \fill[blue!70!black, opacity=0.45] (8.357,0) rectangle (9.286,1.982);
  \fill[blue!70!black, opacity=0.45] (9.286,0) rectangle (10.214,1.261);
  \fill[blue!70!black, opacity=0.45] (10.214,0) rectangle (11.143,1.081);
  \fill[blue!70!black, opacity=0.45] (11.143,0) rectangle (12.071,0.721);
  \fill[blue!70!black, opacity=0.45] (12.071,0) rectangle (13.000,0.360);
  \fill[green!45!black, opacity=0.45] (0.000,0) rectangle (0.929,0.006);
  \fill[green!45!black, opacity=0.45] (2.786,0) rectangle (3.714,0.157);
  \fill[green!45!black, opacity=0.45] (4.643,0) rectangle (5.571,0.012);
  \fill[green!45!black, opacity=0.45] (5.571,0) rectangle (6.500,0.160);
  \fill[green!45!black, opacity=0.45] (6.500,0) rectangle (7.429,6.400);
  \fill[green!45!black, opacity=0.45] (7.429,0) rectangle (8.357,2.389);
  \fill[green!45!black, opacity=0.45] (8.357,0) rectangle (9.286,1.345);
  \fill[green!45!black, opacity=0.45] (9.286,0) rectangle (10.214,1.455);
  \fill[green!45!black, opacity=0.45] (10.214,0) rectangle (11.143,0.040);
  \fill[green!45!black, opacity=0.45] (11.143,0) rectangle (12.071,1.328);
  \fill[green!45!black, opacity=0.45] (12.071,0) rectangle (13.000,0.042);
  \fill[red!75!black] (6.328,0) circle (2.6pt);
  \fill[red!75!black] (7.827,0) circle (2.6pt);
  \fill[red!75!black] (9.549,0) circle (2.6pt);
  \fill[red!75!black] (8.291,0) circle (2.6pt);
  \fill[red!75!black] (9.372,0) circle (2.6pt);
  \fill[red!75!black] (7.683,0) circle (2.6pt);
  \draw[black!70, line width=0.7pt] (0,0) -- (13.000,0);
  \draw[black!70, line width=0.7pt] (0,0) -- (0,6.400);
  \draw[black!70] (0.929,0) -- (0.929,-0.14);
  \node[below, font=\scriptsize] at (0.929,-0.14) {-30};
  \draw[black!70] (2.786,0) -- (2.786,-0.14);
  \node[below, font=\scriptsize] at (2.786,-0.14) {-20};
  \draw[black!70] (4.643,0) -- (4.643,-0.14);
  \node[below, font=\scriptsize] at (4.643,-0.14) {-10};
  \draw[black!70] (6.500,0) -- (6.500,-0.14);
  \node[below, font=\scriptsize] at (6.500,-0.14) {0};
  \draw[black!70] (8.357,0) -- (8.357,-0.14);
  \node[below, font=\scriptsize] at (8.357,-0.14) {10};
  \draw[black!70] (10.214,0) -- (10.214,-0.14);
  \node[below, font=\scriptsize] at (10.214,-0.14) {20};
  \draw[black!70] (12.071,0) -- (12.071,-0.14);
  \node[below, font=\scriptsize] at (12.071,-0.14) {30};
  \draw[black!70] (0,0.000) -- (-0.12,0.000);
  \node[left, font=\scriptsize] at (-0.12,0.000) {0};
  \draw[black!70] (0,1.802) -- (-0.12,1.802);
  \node[left, font=\scriptsize] at (-0.12,1.802) {10};
  \draw[black!70] (0,3.604) -- (-0.12,3.604);
  \node[left, font=\scriptsize] at (-0.12,3.604) {20};
  \draw[black!70] (0,5.406) -- (-0.12,5.406);
  \node[left, font=\scriptsize] at (-0.12,5.406) {30};
  \node[below, font=\small] at (6.500,-0.85) {Profit, Thousands of MAD per Year};
  \node[rotate=90, above, font=\small] at (-1.05,3.200) {Pairs (Weighted)};
  \fill[blue!70!black, opacity=0.45] (7.800,6.100) rectangle (8.250,6.420);
  \node[right, font=\scriptsize] at (8.350,6.260) {Candidate Population $P_X$};
  \fill[green!45!black, opacity=0.45] (7.800,5.500) rectangle (8.250,5.820);
  \node[right, font=\scriptsize] at (8.350,5.660) {Deployment Population $Q_{\theta^*}$};
  \fill[red!75!black] (8.020,5.060) circle (2.6pt);
  \node[right, font=\scriptsize] at (8.350,5.060) {W-DRO Selection};
\end{tikzpicture}
}
\end{figure}

The estimate of the $\ATE(Q)$ under $||\theta|| = 1$ is $-723$ and $-2252$ under $||\theta|| = 2$, indicating that these shifts reverse the sign of the paper's main finding.

\begin{table}[H]\centering
\caption{Simulation design for the Cr\'epon re-analysis.}
\label{tab:crepon_design}
\begin{tabular}{@{}l p{0.78\textwidth}@{}}
\toprule
Step & Description \\
\midrule
1. Sites & The matched village pairs, with site-level effects estimated as the treated--control difference in mean profit within each pair. \\
\addlinespace
2. Candidate set & All pairs, each with equal weight. Site selection methods observe the pair-level baseline covariates. \\
\addlinespace
3. Selection & Each method selects $K = 6$ pairs. The Wasserstein DRO selection hedges over tilts toward low market potential, up to $\rho = \|\theta^*\|$. \\
\addlinespace
4. Deployment set & The same pairs reweighted by $q_s \propto \exp(\theta^{*\prime} X_s)$, with $\theta^*$ pointing toward low market potential at $\|\theta^*\| \in \{0, 1, 2\}$ standard deviations of that index. \\
\addlinespace
5. Estimate & Compare the equal-weight estimate from the selected pairs to the PATE under the deployment weights. \\
\addlinespace
6. Report & Mean absolute bias, mean squared error, and mean $W_1$ distance from the selected sites to the deployment population. Deterministic methods appear once; stratified and random selection average 100 draws. \\
\bottomrule
\end{tabular}
\end{table}

\begin{figure}[H]\centering
\label{fig:crepon_tilt}
\begin{subfigure}{\textwidth}\centering
\resizebox{0.74\textwidth}{!}{%
\begin{tikzpicture}
  \draw[->, gray!65, line width=0.6pt] (-4.00,0) -- (4.42,0) node[right, font=\scriptsize, text=gray!75!black] {market potential (PC$_1$)};
  \draw[->, gray!65, line width=0.6pt] (0,-3.81) -- (0,5.67) node[above, font=\scriptsize, text=gray!75!black] {PC$_2$};
  \fill[blue!70!black, opacity=0.22] (-1.372,-0.989) circle (2.6pt);
  \fill[blue!70!black, opacity=0.22] (0.335,2.140) circle (2.6pt);
  \fill[blue!70!black, opacity=0.22] (1.206,-1.695) circle (2.6pt);
  \fill[blue!70!black, opacity=0.22] (1.428,1.178) circle (2.6pt);
  \fill[blue!70!black, opacity=0.22] (3.108,-0.511) circle (2.6pt);
  \fill[blue!70!black, opacity=0.22] (-1.085,0.040) circle (2.6pt);
  \fill[blue!70!black, opacity=0.22] (3.142,-3.114) circle (2.6pt);
  \fill[blue!70!black, opacity=0.22] (2.926,4.574) circle (2.6pt);
  \fill[blue!70!black, opacity=0.22] (-1.895,-0.296) circle (2.6pt);
  \fill[blue!70!black, opacity=0.22] (0.184,-1.709) circle (2.6pt);
  \fill[blue!70!black, opacity=0.22] (0.877,-2.238) circle (2.6pt);
  \fill[blue!70!black, opacity=0.22] (0.744,-2.236) circle (2.6pt);
  \fill[blue!70!black, opacity=0.22] (-0.088,1.306) circle (2.6pt);
  \fill[blue!70!black, opacity=0.22] (3.416,1.054) circle (2.6pt);
  \fill[blue!70!black, opacity=0.22] (0.473,0.614) circle (2.6pt);
  \fill[blue!70!black, opacity=0.22] (-2.783,0.625) circle (2.6pt);
  \fill[blue!70!black, opacity=0.22] (-0.679,-0.162) circle (2.6pt);
  \fill[blue!70!black, opacity=0.22] (1.688,-1.075) circle (2.6pt);
  \fill[blue!70!black, opacity=0.22] (-3.198,3.056) circle (2.6pt);
  \fill[blue!70!black, opacity=0.22] (0.186,-0.281) circle (2.6pt);
  \fill[blue!70!black, opacity=0.22] (0.963,-0.336) circle (2.6pt);
  \fill[blue!70!black, opacity=0.22] (-0.435,-0.929) circle (2.6pt);
  \fill[blue!70!black, opacity=0.22] (-0.458,2.969) circle (2.6pt);
  \fill[blue!70!black, opacity=0.22] (-0.365,-1.989) circle (2.6pt);
  \fill[blue!70!black, opacity=0.22] (-1.722,-1.040) circle (2.6pt);
  \fill[blue!70!black, opacity=0.22] (-2.450,0.056) circle (2.6pt);
  \fill[blue!70!black, opacity=0.22] (-0.635,2.816) circle (2.6pt);
  \fill[blue!70!black, opacity=0.22] (-2.955,0.527) circle (2.6pt);
  \fill[blue!70!black, opacity=0.22] (1.697,-2.255) circle (2.6pt);
  \fill[blue!70!black, opacity=0.22] (1.568,-1.835) circle (2.6pt);
  \fill[blue!70!black, opacity=0.22] (1.818,1.075) circle (2.6pt);
  \fill[blue!70!black, opacity=0.22] (0.416,-2.284) circle (2.6pt);
  \fill[blue!70!black, opacity=0.22] (1.112,-0.520) circle (2.6pt);
  \fill[blue!70!black, opacity=0.22] (1.698,0.323) circle (2.6pt);
  \fill[blue!70!black, opacity=0.22] (0.449,-0.836) circle (2.6pt);
  \fill[blue!70!black, opacity=0.22] (0.102,-0.277) circle (2.6pt);
  \fill[blue!70!black, opacity=0.22] (-2.031,-0.491) circle (2.6pt);
  \fill[blue!70!black, opacity=0.22] (1.624,-2.312) circle (2.6pt);
  \fill[blue!70!black, opacity=0.22] (0.118,1.609) circle (2.6pt);
  \fill[blue!70!black, opacity=0.22] (-0.795,0.268) circle (2.6pt);
  \fill[blue!70!black, opacity=0.22] (2.770,0.326) circle (2.6pt);
  \fill[blue!70!black, opacity=0.22] (1.462,-2.982) circle (2.6pt);
  \fill[blue!70!black, opacity=0.22] (-0.398,4.870) circle (2.6pt);
  \fill[blue!70!black, opacity=0.22] (-0.794,-1.824) circle (2.6pt);
  \fill[blue!70!black, opacity=0.22] (-1.123,-0.728) circle (2.6pt);
  \fill[blue!70!black, opacity=0.22] (0.951,1.245) circle (2.6pt);
  \fill[blue!70!black, opacity=0.22] (-1.768,0.222) circle (2.6pt);
  \fill[blue!70!black, opacity=0.22] (-0.932,-1.913) circle (2.6pt);
  \fill[blue!70!black, opacity=0.22] (2.957,0.303) circle (2.6pt);
  \fill[blue!70!black, opacity=0.22] (1.336,2.925) circle (2.6pt);
  \fill[blue!70!black, opacity=0.22] (1.370,1.414) circle (2.6pt);
  \fill[blue!70!black, opacity=0.22] (-2.461,0.737) circle (2.6pt);
  \fill[blue!70!black, opacity=0.22] (0.824,-1.389) circle (2.6pt);
  \fill[blue!70!black, opacity=0.22] (-1.417,-1.147) circle (2.6pt);
  \fill[blue!70!black, opacity=0.22] (-1.052,-0.531) circle (2.6pt);
  \fill[blue!70!black, opacity=0.22] (-2.809,0.937) circle (2.6pt);
  \fill[blue!70!black, opacity=0.22] (0.958,0.738) circle (2.6pt);
  \fill[blue!70!black, opacity=0.22] (-2.656,0.327) circle (2.6pt);
  \fill[blue!70!black, opacity=0.22] (0.392,-1.716) circle (2.6pt);
  \fill[blue!70!black, opacity=0.22] (1.591,-1.626) circle (2.6pt);
  \fill[blue!70!black, opacity=0.22] (-1.230,1.552) circle (2.6pt);
  \fill[blue!70!black, opacity=0.22] (1.374,-2.160) circle (2.6pt);
  \fill[blue!70!black, opacity=0.22] (-0.254,-0.164) circle (2.6pt);
  \fill[blue!70!black, opacity=0.22] (-2.081,0.135) circle (2.6pt);
  \fill[blue!70!black, opacity=0.22] (-0.491,-0.009) circle (2.6pt);
  \fill[blue!70!black, opacity=0.22] (-1.909,0.256) circle (2.6pt);
  \fill[blue!70!black, opacity=0.22] (-3.057,1.515) circle (2.6pt);
  \fill[blue!70!black, opacity=0.22] (1.161,3.281) circle (2.6pt);
  \fill[blue!70!black, opacity=0.22] (1.194,0.587) circle (2.6pt);
  \fill[blue!70!black, opacity=0.22] (-2.240,-0.000) circle (2.6pt);
  \fill[green!45!black, opacity=0.80] (-1.372,-0.989) circle (2.6pt);
  \fill[green!45!black, opacity=0.80] (0.335,2.140) circle (0.9pt);
  \fill[green!45!black, opacity=0.80] (1.206,-1.695) circle (0.7pt);
  \fill[green!45!black, opacity=0.80] (1.428,1.178) circle (0.7pt);
  \fill[green!45!black, opacity=0.80] (3.108,-0.511) circle (0.7pt);
  \fill[green!45!black, opacity=0.80] (-1.085,0.040) circle (2.2pt);
  \fill[green!45!black, opacity=0.80] (3.142,-3.114) circle (0.7pt);
  \fill[green!45!black, opacity=0.80] (2.926,4.574) circle (0.7pt);
  \fill[green!45!black, opacity=0.80] (-1.895,-0.296) circle (3.5pt);
  \fill[green!45!black, opacity=0.80] (0.184,-1.709) circle (1.0pt);
  \fill[green!45!black, opacity=0.80] (0.877,-2.238) circle (0.7pt);
  \fill[green!45!black, opacity=0.80] (0.744,-2.236) circle (0.7pt);
  \fill[green!45!black, opacity=0.80] (-0.088,1.306) circle (1.2pt);
  \fill[green!45!black, opacity=0.80] (3.416,1.054) circle (0.7pt);
  \fill[green!45!black, opacity=0.80] (0.473,0.614) circle (0.9pt);
  \fill[green!45!black, opacity=0.80] (-2.783,0.625) circle (5.9pt);
  \fill[green!45!black, opacity=0.80] (-0.679,-0.162) circle (1.7pt);
  \fill[green!45!black, opacity=0.80] (1.688,-1.075) circle (0.7pt);
  \fill[green!45!black, opacity=0.80] (-3.198,3.056) circle (7.5pt);
  \fill[green!45!black, opacity=0.80] (0.186,-0.281) circle (1.0pt);
  \fill[green!45!black, opacity=0.80] (0.963,-0.336) circle (0.7pt);
  \fill[green!45!black, opacity=0.80] (-0.435,-0.929) circle (1.5pt);
  \fill[green!45!black, opacity=0.80] (-0.458,2.969) circle (1.5pt);
  \fill[green!45!black, opacity=0.80] (-0.365,-1.989) circle (1.4pt);
  \fill[green!45!black, opacity=0.80] (-1.722,-1.040) circle (3.2pt);
  \fill[green!45!black, opacity=0.80] (-2.450,0.056) circle (4.9pt);
  \fill[green!45!black, opacity=0.80] (-0.635,2.816) circle (1.7pt);
  \fill[green!45!black, opacity=0.80] (-2.955,0.527) circle (6.6pt);
  \fill[green!45!black, opacity=0.80] (1.697,-2.255) circle (0.7pt);
  \fill[green!45!black, opacity=0.80] (1.568,-1.835) circle (0.7pt);
  \fill[green!45!black, opacity=0.80] (1.818,1.075) circle (0.7pt);
  \fill[green!45!black, opacity=0.80] (0.416,-2.284) circle (0.9pt);
  \fill[green!45!black, opacity=0.80] (1.112,-0.520) circle (0.7pt);
  \fill[green!45!black, opacity=0.80] (1.698,0.323) circle (0.7pt);
  \fill[green!45!black, opacity=0.80] (0.449,-0.836) circle (0.9pt);
  \fill[green!45!black, opacity=0.80] (0.102,-0.277) circle (1.1pt);
  \fill[green!45!black, opacity=0.80] (-2.031,-0.491) circle (3.8pt);
  \fill[green!45!black, opacity=0.80] (1.624,-2.312) circle (0.7pt);
  \fill[green!45!black, opacity=0.80] (0.118,1.609) circle (1.1pt);
  \fill[green!45!black, opacity=0.80] (-0.795,0.268) circle (1.8pt);
  \fill[green!45!black, opacity=0.80] (2.770,0.326) circle (0.7pt);
  \fill[green!45!black, opacity=0.80] (1.462,-2.982) circle (0.7pt);
  \fill[green!45!black, opacity=0.80] (-0.398,4.870) circle (1.5pt);
  \fill[green!45!black, opacity=0.80] (-0.794,-1.824) circle (1.8pt);
  \fill[green!45!black, opacity=0.80] (-1.123,-0.728) circle (2.2pt);
  \fill[green!45!black, opacity=0.80] (0.951,1.245) circle (0.7pt);
  \fill[green!45!black, opacity=0.80] (-1.768,0.222) circle (3.3pt);
  \fill[green!45!black, opacity=0.80] (-0.932,-1.913) circle (2.0pt);
  \fill[green!45!black, opacity=0.80] (2.957,0.303) circle (0.7pt);
  \fill[green!45!black, opacity=0.80] (1.336,2.925) circle (0.7pt);
  \fill[green!45!black, opacity=0.80] (1.370,1.414) circle (0.7pt);
  \fill[green!45!black, opacity=0.80] (-2.461,0.737) circle (4.9pt);
  \fill[green!45!black, opacity=0.80] (0.824,-1.389) circle (0.7pt);
  \fill[green!45!black, opacity=0.80] (-1.417,-1.147) circle (2.7pt);
  \fill[green!45!black, opacity=0.80] (-1.052,-0.531) circle (2.1pt);
  \fill[green!45!black, opacity=0.80] (-2.809,0.937) circle (6.0pt);
  \fill[green!45!black, opacity=0.80] (0.958,0.738) circle (0.7pt);
  \fill[green!45!black, opacity=0.80] (-2.656,0.327) circle (5.5pt);
  \fill[green!45!black, opacity=0.80] (0.392,-1.716) circle (0.9pt);
  \fill[green!45!black, opacity=0.80] (1.591,-1.626) circle (0.7pt);
  \fill[green!45!black, opacity=0.80] (-1.230,1.552) circle (2.4pt);
  \fill[green!45!black, opacity=0.80] (1.374,-2.160) circle (0.7pt);
  \fill[green!45!black, opacity=0.80] (-0.254,-0.164) circle (1.3pt);
  \fill[green!45!black, opacity=0.80] (-2.081,0.135) circle (3.9pt);
  \fill[green!45!black, opacity=0.80] (-0.491,-0.009) circle (1.5pt);
  \fill[green!45!black, opacity=0.80] (-1.909,0.256) circle (3.5pt);
  \fill[green!45!black, opacity=0.80] (-3.057,1.515) circle (7.0pt);
  \fill[green!45!black, opacity=0.80] (1.161,3.281) circle (0.7pt);
  \fill[green!45!black, opacity=0.80] (1.194,0.587) circle (0.7pt);
  \fill[green!45!black, opacity=0.80] (-2.240,-0.000) circle (4.3pt);
  \draw[->, green!35!black, line width=1.6pt] (0.000,-0.000) -- (-2.310,0.704);
  \node[font=\small, green!30!black] at (-1.155,0.902) {$\theta^*$};
  \draw[green!35!black, line width=1.5pt] (-2.310,0.704) ++(-5pt,-5pt) -- ++(10pt,10pt);
  \draw[green!35!black, line width=1.5pt] (-2.310,0.704) ++(-5pt,5pt) -- ++(10pt,-10pt);
\end{tikzpicture}
}
\caption{Tilted sites in the direction of low market potential}
\label{fig:crepon_tilt_a}
\end{subfigure}

\begin{subfigure}{\textwidth}\centering
\resizebox{0.74\textwidth}{!}{%
\begin{tikzpicture}
  \draw[->, gray!65, line width=0.6pt] (-4.00,0) -- (4.42,0) node[right, font=\scriptsize, text=gray!75!black] {market potential (PC$_1$)};
  \draw[->, gray!65, line width=0.6pt] (0,-3.81) -- (0,5.67) node[above, font=\scriptsize, text=gray!75!black] {PC$_2$};
  \fill[green!45!black, opacity=0.32] (-1.372,-0.989) circle (2.6pt);
  \fill[green!45!black, opacity=0.32] (0.335,2.140) circle (0.9pt);
  \fill[green!45!black, opacity=0.32] (1.206,-1.695) circle (0.7pt);
  \fill[green!45!black, opacity=0.32] (1.428,1.178) circle (0.7pt);
  \fill[green!45!black, opacity=0.32] (3.108,-0.511) circle (0.7pt);
  \fill[green!45!black, opacity=0.32] (-1.085,0.040) circle (2.2pt);
  \fill[green!45!black, opacity=0.32] (3.142,-3.114) circle (0.7pt);
  \fill[green!45!black, opacity=0.32] (2.926,4.574) circle (0.7pt);
  \fill[green!45!black, opacity=0.32] (-1.895,-0.296) circle (3.5pt);
  \fill[green!45!black, opacity=0.32] (0.184,-1.709) circle (1.0pt);
  \fill[green!45!black, opacity=0.32] (0.877,-2.238) circle (0.7pt);
  \fill[green!45!black, opacity=0.32] (0.744,-2.236) circle (0.7pt);
  \fill[green!45!black, opacity=0.32] (-0.088,1.306) circle (1.2pt);
  \fill[green!45!black, opacity=0.32] (3.416,1.054) circle (0.7pt);
  \fill[green!45!black, opacity=0.32] (0.473,0.614) circle (0.9pt);
  \fill[green!45!black, opacity=0.32] (-2.783,0.625) circle (5.9pt);
  \fill[green!45!black, opacity=0.32] (-0.679,-0.162) circle (1.7pt);
  \fill[green!45!black, opacity=0.32] (1.688,-1.075) circle (0.7pt);
  \fill[green!45!black, opacity=0.32] (-3.198,3.056) circle (7.5pt);
  \fill[green!45!black, opacity=0.32] (0.186,-0.281) circle (1.0pt);
  \fill[green!45!black, opacity=0.32] (0.963,-0.336) circle (0.7pt);
  \fill[green!45!black, opacity=0.32] (-0.435,-0.929) circle (1.5pt);
  \fill[green!45!black, opacity=0.32] (-0.458,2.969) circle (1.5pt);
  \fill[green!45!black, opacity=0.32] (-0.365,-1.989) circle (1.4pt);
  \fill[green!45!black, opacity=0.32] (-1.722,-1.040) circle (3.2pt);
  \fill[green!45!black, opacity=0.32] (-2.450,0.056) circle (4.9pt);
  \fill[green!45!black, opacity=0.32] (-0.635,2.816) circle (1.7pt);
  \fill[green!45!black, opacity=0.32] (-2.955,0.527) circle (6.6pt);
  \fill[green!45!black, opacity=0.32] (1.697,-2.255) circle (0.7pt);
  \fill[green!45!black, opacity=0.32] (1.568,-1.835) circle (0.7pt);
  \fill[green!45!black, opacity=0.32] (1.818,1.075) circle (0.7pt);
  \fill[green!45!black, opacity=0.32] (0.416,-2.284) circle (0.9pt);
  \fill[green!45!black, opacity=0.32] (1.112,-0.520) circle (0.7pt);
  \fill[green!45!black, opacity=0.32] (1.698,0.323) circle (0.7pt);
  \fill[green!45!black, opacity=0.32] (0.449,-0.836) circle (0.9pt);
  \fill[green!45!black, opacity=0.32] (0.102,-0.277) circle (1.1pt);
  \fill[green!45!black, opacity=0.32] (-2.031,-0.491) circle (3.8pt);
  \fill[green!45!black, opacity=0.32] (1.624,-2.312) circle (0.7pt);
  \fill[green!45!black, opacity=0.32] (0.118,1.609) circle (1.1pt);
  \fill[green!45!black, opacity=0.32] (-0.795,0.268) circle (1.8pt);
  \fill[green!45!black, opacity=0.32] (2.770,0.326) circle (0.7pt);
  \fill[green!45!black, opacity=0.32] (1.462,-2.982) circle (0.7pt);
  \fill[green!45!black, opacity=0.32] (-0.398,4.870) circle (1.5pt);
  \fill[green!45!black, opacity=0.32] (-0.794,-1.824) circle (1.8pt);
  \fill[green!45!black, opacity=0.32] (-1.123,-0.728) circle (2.2pt);
  \fill[green!45!black, opacity=0.32] (0.951,1.245) circle (0.7pt);
  \fill[green!45!black, opacity=0.32] (-1.768,0.222) circle (3.3pt);
  \fill[green!45!black, opacity=0.32] (-0.932,-1.913) circle (2.0pt);
  \fill[green!45!black, opacity=0.32] (2.957,0.303) circle (0.7pt);
  \fill[green!45!black, opacity=0.32] (1.336,2.925) circle (0.7pt);
  \fill[green!45!black, opacity=0.32] (1.370,1.414) circle (0.7pt);
  \fill[green!45!black, opacity=0.32] (-2.461,0.737) circle (4.9pt);
  \fill[green!45!black, opacity=0.32] (0.824,-1.389) circle (0.7pt);
  \fill[green!45!black, opacity=0.32] (-1.417,-1.147) circle (2.7pt);
  \fill[green!45!black, opacity=0.32] (-1.052,-0.531) circle (2.1pt);
  \fill[green!45!black, opacity=0.32] (-2.809,0.937) circle (6.0pt);
  \fill[green!45!black, opacity=0.32] (0.958,0.738) circle (0.7pt);
  \fill[green!45!black, opacity=0.32] (-2.656,0.327) circle (5.5pt);
  \fill[green!45!black, opacity=0.32] (0.392,-1.716) circle (0.9pt);
  \fill[green!45!black, opacity=0.32] (1.591,-1.626) circle (0.7pt);
  \fill[green!45!black, opacity=0.32] (-1.230,1.552) circle (2.4pt);
  \fill[green!45!black, opacity=0.32] (1.374,-2.160) circle (0.7pt);
  \fill[green!45!black, opacity=0.32] (-0.254,-0.164) circle (1.3pt);
  \fill[green!45!black, opacity=0.32] (-2.081,0.135) circle (3.9pt);
  \fill[green!45!black, opacity=0.32] (-0.491,-0.009) circle (1.5pt);
  \fill[green!45!black, opacity=0.32] (-1.909,0.256) circle (3.5pt);
  \fill[green!45!black, opacity=0.32] (-3.057,1.515) circle (7.0pt);
  \fill[green!45!black, opacity=0.32] (1.161,3.281) circle (0.7pt);
  \fill[green!45!black, opacity=0.32] (1.194,0.587) circle (0.7pt);
  \fill[green!45!black, opacity=0.32] (-2.240,-0.000) circle (4.3pt);
  \fill[blue!70!black, opacity=0.45] (-1.372,-0.989) circle (2.4pt);
  \fill[blue!70!black, opacity=0.45] (0.335,2.140) circle (2.4pt);
  \fill[blue!70!black, opacity=0.45] (1.206,-1.695) circle (2.4pt);
  \fill[blue!70!black, opacity=0.45] (1.428,1.178) circle (2.4pt);
  \fill[blue!70!black, opacity=0.45] (3.108,-0.511) circle (2.4pt);
  \fill[blue!70!black, opacity=0.45] (-1.085,0.040) circle (2.4pt);
  \fill[blue!70!black, opacity=0.45] (3.142,-3.114) circle (2.4pt);
  \fill[blue!70!black, opacity=0.45] (2.926,4.574) circle (2.4pt);
  \fill[blue!70!black, opacity=0.45] (-1.895,-0.296) circle (2.4pt);
  \fill[blue!70!black, opacity=0.45] (0.184,-1.709) circle (2.4pt);
  \fill[blue!70!black, opacity=0.45] (0.877,-2.238) circle (2.4pt);
  \fill[blue!70!black, opacity=0.45] (0.744,-2.236) circle (2.4pt);
  \fill[blue!70!black, opacity=0.45] (-0.088,1.306) circle (2.4pt);
  \fill[blue!70!black, opacity=0.45] (3.416,1.054) circle (2.4pt);
  \fill[blue!70!black, opacity=0.45] (0.473,0.614) circle (2.4pt);
  \fill[blue!70!black, opacity=0.45] (-2.783,0.625) circle (2.4pt);
  \fill[blue!70!black, opacity=0.45] (-0.679,-0.162) circle (2.4pt);
  \fill[blue!70!black, opacity=0.45] (1.688,-1.075) circle (2.4pt);
  \fill[blue!70!black, opacity=0.45] (-3.198,3.056) circle (2.4pt);
  \fill[blue!70!black, opacity=0.45] (0.186,-0.281) circle (2.4pt);
  \fill[blue!70!black, opacity=0.45] (0.963,-0.336) circle (2.4pt);
  \fill[blue!70!black, opacity=0.45] (-0.435,-0.929) circle (2.4pt);
  \fill[blue!70!black, opacity=0.45] (-0.458,2.969) circle (2.4pt);
  \fill[blue!70!black, opacity=0.45] (-0.365,-1.989) circle (2.4pt);
  \fill[blue!70!black, opacity=0.45] (-1.722,-1.040) circle (2.4pt);
  \fill[blue!70!black, opacity=0.45] (-2.450,0.056) circle (2.4pt);
  \fill[blue!70!black, opacity=0.45] (-0.635,2.816) circle (2.4pt);
  \fill[blue!70!black, opacity=0.45] (-2.955,0.527) circle (2.4pt);
  \fill[blue!70!black, opacity=0.45] (1.697,-2.255) circle (2.4pt);
  \fill[blue!70!black, opacity=0.45] (1.568,-1.835) circle (2.4pt);
  \fill[blue!70!black, opacity=0.45] (1.818,1.075) circle (2.4pt);
  \fill[blue!70!black, opacity=0.45] (0.416,-2.284) circle (2.4pt);
  \fill[blue!70!black, opacity=0.45] (1.112,-0.520) circle (2.4pt);
  \fill[blue!70!black, opacity=0.45] (1.698,0.323) circle (2.4pt);
  \fill[blue!70!black, opacity=0.45] (0.449,-0.836) circle (2.4pt);
  \fill[blue!70!black, opacity=0.45] (0.102,-0.277) circle (2.4pt);
  \fill[blue!70!black, opacity=0.45] (-2.031,-0.491) circle (2.4pt);
  \fill[blue!70!black, opacity=0.45] (1.624,-2.312) circle (2.4pt);
  \fill[blue!70!black, opacity=0.45] (0.118,1.609) circle (2.4pt);
  \fill[blue!70!black, opacity=0.45] (-0.795,0.268) circle (2.4pt);
  \fill[blue!70!black, opacity=0.45] (2.770,0.326) circle (2.4pt);
  \fill[blue!70!black, opacity=0.45] (1.462,-2.982) circle (2.4pt);
  \fill[blue!70!black, opacity=0.45] (-0.398,4.870) circle (2.4pt);
  \fill[blue!70!black, opacity=0.45] (-0.794,-1.824) circle (2.4pt);
  \fill[blue!70!black, opacity=0.45] (-1.123,-0.728) circle (2.4pt);
  \fill[blue!70!black, opacity=0.45] (0.951,1.245) circle (2.4pt);
  \fill[blue!70!black, opacity=0.45] (-1.768,0.222) circle (2.4pt);
  \fill[blue!70!black, opacity=0.45] (-0.932,-1.913) circle (2.4pt);
  \fill[blue!70!black, opacity=0.45] (2.957,0.303) circle (2.4pt);
  \fill[blue!70!black, opacity=0.45] (1.336,2.925) circle (2.4pt);
  \fill[blue!70!black, opacity=0.45] (1.370,1.414) circle (2.4pt);
  \fill[blue!70!black, opacity=0.45] (-2.461,0.737) circle (2.4pt);
  \fill[blue!70!black, opacity=0.45] (0.824,-1.389) circle (2.4pt);
  \fill[blue!70!black, opacity=0.45] (-1.417,-1.147) circle (2.4pt);
  \fill[blue!70!black, opacity=0.45] (-1.052,-0.531) circle (2.4pt);
  \fill[blue!70!black, opacity=0.45] (-2.809,0.937) circle (2.4pt);
  \fill[blue!70!black, opacity=0.45] (0.958,0.738) circle (2.4pt);
  \fill[blue!70!black, opacity=0.45] (-2.656,0.327) circle (2.4pt);
  \fill[blue!70!black, opacity=0.45] (0.392,-1.716) circle (2.4pt);
  \fill[blue!70!black, opacity=0.45] (1.591,-1.626) circle (2.4pt);
  \fill[blue!70!black, opacity=0.45] (-1.230,1.552) circle (2.4pt);
  \fill[blue!70!black, opacity=0.45] (1.374,-2.160) circle (2.4pt);
  \fill[blue!70!black, opacity=0.45] (-0.254,-0.164) circle (2.4pt);
  \fill[blue!70!black, opacity=0.45] (-2.081,0.135) circle (2.4pt);
  \fill[blue!70!black, opacity=0.45] (-0.491,-0.009) circle (2.4pt);
  \fill[blue!70!black, opacity=0.45] (-1.909,0.256) circle (2.4pt);
  \fill[blue!70!black, opacity=0.45] (-3.057,1.515) circle (2.4pt);
  \fill[blue!70!black, opacity=0.45] (1.161,3.281) circle (2.4pt);
  \fill[blue!70!black, opacity=0.45] (1.194,0.587) circle (2.4pt);
  \fill[blue!70!black, opacity=0.45] (-2.240,-0.000) circle (2.4pt);
  \fill[red!75!black, opacity=0.95] (-1.085,0.040) circle (4.4pt);
  \fill[red!75!black, opacity=0.95] (-2.783,0.625) circle (4.4pt);
  \fill[red!75!black, opacity=0.95] (-0.365,-1.989) circle (4.4pt);
  \fill[red!75!black, opacity=0.95] (-2.450,0.056) circle (4.4pt);
  \fill[red!75!black, opacity=0.95] (-0.635,2.816) circle (4.4pt);
  \fill[red!75!black, opacity=0.95] (1.112,-0.520) circle (4.4pt);
  \draw[green!35!black, line width=1.5pt] (-2.310,0.704) ++(-5pt,-5pt) -- ++(10pt,10pt);
  \draw[green!35!black, line width=1.5pt] (-2.310,0.704) ++(-5pt,5pt) -- ++(10pt,-10pt);
  \draw[red!75!black, line width=1.5pt] (-1.034,0.171) ++(-5pt,-5pt) -- ++(10pt,10pt);
  \draw[red!75!black, line width=1.5pt] (-1.034,0.171) ++(-5pt,5pt) -- ++(10pt,-10pt);
\end{tikzpicture}
}
\caption{The Wasserstein DRO selection (red), with the ATE(S), ATE(Q) indicated.}
\label{fig:crepon_tilt_b}
\end{subfigure}
\caption{The Cr\'epon design under a tilt toward low market potential, $\|\theta^*\| = 2$.}
\end{figure}

\subsubsection*{Results}

\begin{table}[H]\centering
\caption{Site selection under a tilt toward low market potential: self-employment profit.}
\label{tab:crepon}

\begin{subtable}{\textwidth}\centering
\caption{Mean absolute bias, dirhams per year.}
\label{tab:crepon_bias}
\begin{tabular}{lrrr}
\toprule
method & $\|\theta^*\| = 0$ & $1$ & $2$ \\
\midrule
W-DRO & \textbf{82} & \textbf{2181} & \textbf{1660} \\
Olea et al. & 84 & 2517 & 3611 \\
Egami et al. & 244 & 2677 & 3771 \\
Tipton & 2073 & 4506 & 5599 \\
Stratified & 1239 & 3445 & 4538 \\
Random & 1332 & 2643 & 3716 \\
\bottomrule
\end{tabular}
\end{subtable}

\vspace{1.2em}
\begin{subtable}{\textwidth}\centering
\caption{Mean squared error, millions of squared dirhams.}
\label{tab:crepon_mse}
\begin{tabular}{lrrr}
\toprule
method & $\|\theta^*\| = 0$ & $1$ & $2$ \\
\midrule
W-DRO & \textbf{0.01} & \textbf{4.76} & \textbf{2.76} \\
Olea et al. & 0.01 & 6.34 & 13.04 \\
Egami et al. & 0.06 & 7.17 & 14.22 \\
Tipton & 4.30 & 20.30 & 31.35 \\
Stratified & 2.31 & 13.15 & 21.88 \\
Random & 2.53 & 9.37 & 16.30 \\
\bottomrule
\end{tabular}
\end{subtable}

\vspace{1.2em}
\begin{subtable}{\textwidth}\centering
\caption{Mean $W_1$ distance to the deployment population, standard-deviation units.}
\label{tab:crepon_w1}
\begin{tabular}{lrrr}
\toprule
method & $\|\theta^*\| = 0$ & $1$ & $2$ \\
\midrule
W-DRO & 1.068 & \textbf{1.120} & \textbf{1.219} \\
Olea et al. & \textbf{0.994} & 1.322 & 1.714 \\
Egami et al. & 2.275 & 2.663 & 3.015 \\
Tipton & 1.252 & 1.676 & 2.066 \\
Stratified & 1.698 & 2.118 & 2.492 \\
Random & 1.494 & 1.639 & 1.939 \\
\bottomrule
\end{tabular}
\end{subtable}
\end{table}

\subsection{Discussion}
In each case, the WDRO method described above performs best under distribution shift under a specified direction. This is because it is design to protect against the particular adversarial case constructed in the simulation example. This is intended to validate the method: when deployment risks are known, the WDRO method effectively hedges against those risks, in terms of minimizing the expected error of downstream estimands as a function of unobserved drift. 

Several qualifications are worth making. First, these simulations presuppose that the direction of the shift is known in advance. This is unrealistic in practice. But the purpose of using these methods is to generate designs that are conservative with respect to specified threats to external validity. Thus, using the method gives you a guarantee of conservatism against possible threats: that is the nature of the statistical guarantee, and is valuable for a study designer who has not yet run a study. Second, there is a clear role for substantive knowledge to guide the generation of potential threats. This creates researcher degrees of freedom, but the adjustment is nonetheless conservative, as it allows the researcher to pre-empt possible threats to external validity and guard against them. 

A third point is that the other methods assessed here are not directional: they hedge in every possible direction simultaneously. This means that they conservatively protect against many possible directions of shift. In practice, researchers are likely to believe that some directions are worse than others in terms of the significance of results.

\section{Conclusions}\label{sec:conclusions}
I develop a novel method for site selection that provides guarantees against directional distribution shifts during the planning stage. This method is useful for practitioners in applied contexts designing multi-site studies when specific threats to external validity are known. This method then allows researchers to choose sites with minimax guarantees against specific threats to external validity.

This method is intended to address the existence of well-documented threats to the validity of large-scale policy evaluations, where site selection bias threatens to lead to biased or inaccurate conclusions drawn from randomized control trials \citep{Allcott_2015, Olsen2022}. 

This method could be encouraged by funding bodies with specific objectives in terms of ensuring representation of minority groups in particular trials. 

The utility of optimization-based methods in general requires that covariates are collected at the planning stage, and are informative.  How can researchers assess the relationship between covariates and expected effects? They can refer to prior experiments; assess the relationship between covariates and pre-treatment outcomes; refer to estimated effects from pilot studies or studies in the same substantive domain; and use domain expertise and substantive knowledge to make suppositions grounded in a mechanistic understanding of the causal relationship. 

Two rhetorical aims of this project is to encourage funding bodies that support multi-site studies to 1) explicitly state threats to external validity ex ante, and commit to experimental designs that pre-empt these threats 2) to commit to collecting informative covariates about trial populations in advance of conducting studies. 

In future work, I intend to extend this project to optimal enrollment of individuals for estimation of the CATE under distribution shift, and to impose structure on $\theta$ to allow for unit enrollment decisions with guarantees on the Conditional Value-at-Risk of an intervention on the welfare of specified subgroups. 

\printbibliography

\appendix

\section{Proofs}\label{app:theory}
\providecommand{\prop}[1]{\textcolor{red!65!black}{#1}}
\providecommand{\cut}[1]{\textcolor{gray}{\sout{#1}}}
\providecommand{\fn}[1]{\textcolor{blue}{[#1]}}

\subsection*{Preliminaries}

We observe $|P|$ candidate sites indexed by $s$, and want to select a subset of size $K$. Each site has a population size of $N_s$. We write $p_s = N_s / \sum_{r} N_r$ for the share of all individuals who are in site $s$, where $p_s > 0$ for each $s$ and $\sum_s p_s = 1$. In each site, units are sampled into the experiment according to a sampling indicator $W_s \sim \Phi_s$ an experiment is conducted, and individuals are randomized into treatmen treatment $Z_s$ according to a design $\Psi_s$. Each site has observed features $X_s \in \mathbb{R}^d$, and unobserved features $U_s \in \mathbb{R}^k$. 

$P_{X,U} = P$ is the joint population distribution on $\mathbb{R}^d \times \mathbb{R}^k$ with
mass $p_s$ at the point $(X_s, U_s)$, for each site $s$. $P_X$, the observed covariate distribution, places mass $p_s$ at each site-level covariate vector $X_s$. $P_U$ is the distribution of unobservables, and $P_{U|X}$ is the population distribution of unobservables with observables fixed at $X = x$. $S_{X,U}$, $S_X$, $S_U$, $S_{U|X}$ are defined analogously. For a deployment population $Q$, $Q_{X,U}$, $Q_X$, $Q_U$, $Q_{U|X}$ are defined in the same way.

Individual $i$ in site $s$ has treatment effect $\tau_{is}$. We write $\tau_s = \frac{1}{N_s}\sum_{i \in s} \tau_{is}$ for the average treatment effect among the $N_s$ units in site $s$.  

Let $\tau: \mathbb{R}^d \times \mathbb{R}^k \to \mathbb{R}$ be the treatment effect function, so that $\tau_s = \tau(X_s, U_s)$. Then, for a distribution $F$ over $(X,U)$, define:
\[
\mathrm{ATE}(F) \;=\; \int \tau(X,U)  \, dF(X,U) .
\]
Evaluated over the observed candidate population: \[\mathrm{ATE}(P) = \sum_s p_s \tau_s\] This is the average effect over individuals in the population. This is the PATE, but it is helpful to make the dependence on the set $P$ explicit.  

Given a selected subset $S$, define the average treatment effect over the selected sites: 
\[\mathrm{ATE}(S) = \frac{1}{K}\sum_{s \in S} \tau_s\]
This is the Sample Average Treatment Effect (SATE) evaluated on the selected subset $S$. Since $S$ is the choice variable, it is again helpful to write this as a function that ranges over $S$. 

For a hypothetical deployment population $Q$, we likewise define the estimand of interest: \[\mathrm{ATE}(Q) = \int \tau(X,U)  \, dQ(X,U)\]

Let $\omega_s$ be a vector of weights on the selected sites, and $\hat{\tau}_s$ the plug-in estimate of $\tau_s$ from the experiment run in site $s$. Then: 
\[
\hat\tau(S) \;=\; \sum_{s \in S} \omega_s \, \hat{\tau}_s.
\]

\begin{remark}[Balanced transport]
Throughout, $\omega_s = 1/K$, so that $\hat\tau(S)$ is the average of the site-level estimates and $\mathbb{E}[\hat\tau(S)] = \mathrm{ATE}(S)$. Each selected site stands in for an equal share of the population.
\end{remark}

Recall the definition of the 1-Wasserstein distance above. For distributions $\mu, \nu$ on $\R^d$ (or on $\R^k$) with finite first moments, let $\Pi(\mu,\nu)$ be the set of couplings of $\mu$ and $\nu$, and
\[
W_1(\mu, \nu) = \inf_{\gamma \in \Pi(\mu,\nu)} \int \lVert x - x' \rVert \, d\gamma(x, x') .
\]
For the joint space $(X,U)$ with support $\R^d \times \R^k$ define the distance metric:
\begin{equation*}
d\big((x,u),(x',u')\big) = \lVert x - x' \rVert + \lVert u - u' \rVert ,
\end{equation*}
and, for joint distribution $F, G$ on $\R^d \times \R^k$ with finite first moments,
\begin{equation*}
W_1^{X,U}(F, G) = \inf_{\gamma \in\Pi(F, G)} \int d\big((x,u),(x',u')\big)\, d\gamma\big((x,u),(x',u')\big) .
\end{equation*}
This is the 1-Wasserstein distance between two distributions $F$ and $G$. This represents the distance  between $F$ and $G$ under the optimal transport plan from $F$ to $G$ in $(X,U)$-space. 
Every distribution in this appendix is supported on finitely many sites, so each integral is a finite sum and each transport plan is a matrix.

\subsection*{A.1 Proof of \Cref{lem:unbiased}}
This is a familiar result. 
\begin{proof}
\begin{align*}
\E\big[\widehat{ATE}(S)\big]
&= \E\left[\frac1K \sum_{s \in S} \hat\tau_s\right]
    \\
&= \frac1K \sum_{s \in S} \E[\hat\tau_s]
    \\
&= \frac1K \sum_{s \in S} \E_{Z_s \sim \Psi_s,\, W_s \sim \Phi_s}[\hat\tau_s]
   \\
&= \frac1K \sum_{s \in S} \tau(X_s, U_s)
    \\
&= \int \tau(x,u)\, dS_{X,U}(x,u)
   \\
&= ATE(S)
   \qedhere
\end{align*}
\end{proof}

\subsection*{A.2 Proof of \Cref{lem:biasvar}}

\begin{proof}

\begin{align*}
\MSE_Q(S)
&= \E\Big[\big(\ATE(Q) - \widehat{\ATE}(S)\big)^2\Big]
    \\
&= \E\Big[\big( [\ATE(Q) - \ATE(S)] - [\widehat{\ATE}(S) - \ATE(S)] \big)^2\Big]
   \\
&= \big(\ATE(Q) - \ATE(S)\big)^2
   - 2\big(\ATE(Q) - \ATE(S)\big)\,\E\big[\widehat{\ATE}(S) - \ATE(S)\big] \\
&\qquad + \E\Big[\big(\widehat{\ATE}(S) - \ATE(S)\big)^2\Big]
    \\
&= \big(\ATE(Q) - \ATE(S)\big)^2 + \E\Big[\big(\widehat{\ATE}(S) - \ATE(S)\big)^2\Big] \quad\quad [\textrm{By \Cref{lem:unbiased}}]
   \\
&= \big(\ATE(Q) - \ATE(S)\big)^2 + \sigma^2_S
   \qedhere
\end{align*}
\end{proof}

\subsection*{A.3 A Lipschitz--transport bound}

\begin{lemma}[Lipschitz--transport bound]\label{lem:LT}
Let $\tau$ be an $L$-Lipschitz function. Then, under \Cref{asm_Lipschitz}:

\begin{equation*}
\Big| \int \tau\, dF - \int \tau\, dG \Big| \;\le\; L \, W_1^{X,U}(F, G) .
\end{equation*}
\end{lemma}

\begin{proof}
Consider a transport plan $\gamma \in \Gamma(F, G)$.
\begin{align*}
\int \tau\, dF - \int \tau\, dG
&= \int \big[\tau(x,u) - \tau(x',u')\big]\, d\gamma\big((x,u),(x',u')\big)
   \\
\Big| \int \tau\, dF - \int \tau\, dG \Big|
&\le \int \big|\tau(x,u) - \tau(x',u')\big|\, d\gamma
    \\
&\le L \int \big( \lVert x - x' \rVert + \lVert u - u' \rVert \big)\, d\gamma
   && [\textrm{By \Cref{asm_Lipschitz}}] \\
&= L \int d\big((x,u),(x',u')\big)\, d\gamma
\end{align*}
This establishes that:

\[ \Big| \int \tau\, dF - \int \tau\, dG \Big| \leq  L \int d\big((x,u),(x',u')\big)\, d\gamma\]

But since $\gamma$ was an arbitrary transport plan, it follows that this inequality holds under the \textit{optimal} transport plan: 

\[ \Big| \int \tau\, dF - \int \tau\, dG \Big| \leq \underset{\gamma}{\inf} L \int d\big((x,u),(x',u')\big)\, d\gamma\]

So that:

\[ \Big| \int \tau\, dF - \int \tau\, dG \Big| \leq LW_1^{X,U}(F,G) \]
\end{proof}

In words, this establishes that the absolute distance between a Lipschitz function $\tau(X,U)$ evaluate on two distributions $F$ and $G$ is upper bounded by a Lipschitz constant times the cost of the optimal transport plan from $F$ to $G$.

\subsection*{A.4 Proof of \Cref{thm:oracle_Q}}

\begin{proof}
\begin{align*}
\MSE_Q(S)
&= \big(\ATE(Q) - \ATE(S)\big)^2 + \sigma^2_S
   \quad\quad [\textrm{By \Cref{lem:biasvar}}] \\
\ATE(Q) - \ATE(S)
&= \int \tau(x)\, dQ_X(x) - \int \tau(x)\, dS_X(x)
   \\
\big|\ATE(Q) - \ATE(S)\big|
&\le  L\, W_1(Q_X, S_X)
   \quad\quad [\textrm{By \Cref{lem:LT}}] \\
\MSE_Q(S)
&\le L^2\, W_1(Q_X, S_X)^2 + \sigma^2_S
   \qedhere
\end{align*}
\end{proof}

\subsection*{A.5 Bound on unobservables}

\begin{lemma}[Bounding the unobservables]\label{lem:glue}
Let $\gamma^{*} \in \Pi(S_X, Q_X)$ be an optimal coupling of $S_X$ and $Q_X$, with first
coordinate $x \sim S_X$ and second coordinate $x' \sim Q_X$. Define
\begin{equation*}
\eta_{\gamma^{*}}(S, Q) \;=\; \int W_1\big( S_{U \mid X = x}, \, Q_{U \mid X = x'} \big) \, d\gamma^{*}(x, x') ,
\end{equation*}
the average distance between the conditional distributions of the unobservables, given the optimal transport plan $\gamma^*$. Then:
\begin{equation*}
W_1^{X,U}(S_{X,U}, Q_{X,U}) \;\le\; W_1(S_X, Q_X) + \eta_{\gamma^{*}}(S, Q) 
\end{equation*}
\end{lemma}

\begin{proof}
For each pair $(x, x')$ in the support of $\gamma^{*}$, let $\kappa_{x,x'}$ be a transport plan from $Q_{U|X = x'}$ to $S_{U|X=x}$ that attains
$W_1(S_{U \mid X = x}, Q_{U \mid X = x'})$. 

Define a measure $\hat\Gamma$ on
$(\R^d \times \R^k) \times (\R^d \times \R^k)$ by
\begin{equation*}
\hat\Gamma(dx, du, dx', du') = \gamma^{*}(dx, dx')\, \kappa_{x,x'}(du, du')
\end{equation*}

First, we show that $\hat{\Gamma}$ is a coupling of $Q_{X,U}$ and $S_{X,U}$. We first show that integrating over $\hat{\Gamma}$ yields these marginals. 

First, integrating out $(x, u)$, we get the source distribution:
\begin{align*}
\int_{x, u} \hat\Gamma(dx, du, dx', du')
&= \int_{x} \gamma^{*}(dx, dx') \int_{u} \kappa_{x,x'}(du, du')
    \\
&= \int_{x} \gamma^{*}(dx, dx')\, Q_{U \mid X = x'}(du')
   \\
&= Q_{U \mid X = x'}(du') \int_{x} \gamma^{*}(dx, dx')
    \\
&= Q_{U \mid X = x'}(du')\, Q_X(dx')
   \\
&= Q_{X,U}(dx', du')
\end{align*}

Second, integrating out $(x',u')$, we get the target distribution: 
\begin{align*}
\int_{x', u'} \hat\Gamma(dx, du, dx', du')
&= \int_{x'} \gamma^{*}(dx, dx') \int_{u'} \kappa_{x,x'}(du, du')
   \\
&= \int_{x'} \gamma^{*}(dx, dx')\, S_{U \mid X = x}(du)
   \\
&= S_X(dx)\, S_{U \mid X = x}(du)
    \\
&= S_{X,U}(dx, du)
\end{align*}

So $\hat\Gamma$ has first marginal $S_{X,U}$ and second marginal $Q_{X,U}$: it is a coupling of the two, and so a transport plan between them.

Next, we show that the cost of the transport plan $\hat{\Gamma}$ is the cost of the optimal transport plan from $Q_X$ to $S_X$ plus the distance from the unobservables of $Q$ to the unobservables of $S$ under the optimal transport plan from $Q_X$ to $S_X$.

\begin{align*}
\int d\, d\hat\Gamma
&= \int \big( \lVert x - x' \rVert + \lVert u - u' \rVert \big)\,
   \gamma^{*}(dx,dx')\, \kappa_{x,x'}(du,du')
    \\
&= \int \lVert x - x' \rVert\, d\gamma^{*}(x,x')
   + \int \Big[ \int \lVert u - u' \rVert\, d\kappa_{x,x'}(u,u') \Big] d\gamma^{*}(x,x')
  \\
&= W_1(S_X, Q_X) + \int W_1\big(S_{U \mid X = x}, Q_{U \mid X = x'}\big)\, d\gamma^{*}(x,x')
    \\
&= W_1(S_X, Q_X) + \eta(S_U, Q_U)
\end{align*}

By definition of the Wasserstein distance under the joint metric $d$,
\[
W_1^{X,U}\big(S_{X,U}, Q_{X,U}\big)
\;=\; \inf_{\Gamma \in \Pi(S_{X,U},\, Q_{X,U})}
\int d\big((x,u),(x',u')\big)\, d\Gamma(x,u,x',u') ,
\]
where $\Pi(S_{X,U}, Q_{X,U})$ is the set of all couplings of $S_{X,U}$ and
$Q_{X,U}$.  Therefore
\begin{align*}
W_1^{X,U}\big(S_{X,U}, Q_{X,U}\big)
&= \inf_{\Gamma \in \Pi(S_{X,U},\, Q_{X,U})} \int d \, d\Gamma
  \le \int d \, d\hat\Gamma
\end{align*}

So that:

\[W_1^{X,U}\big(S_{X,U}, Q_{X,U}\big) \leq W_1(S_X, Q_X) + \eta(S_U, Q_U)\]

\end{proof}

\subsection*{A.6 Proof of \Cref{thm:eta_Q}}

\begin{proof}
\begin{align*}
\MSE_Q(S)
&= \big(\ATE(Q) - \ATE(S)\big)^2 + \sigma^2_S
   && [\textrm{By \Cref{lem:biasvar}}] \\
\big|ATE(Q) - ATE(S)\big|
&= \Big| \int \tau\, dQ_{X,U} - \int \tau\, dS_{X,U} \Big|
   &&  [\textrm{By \Cref{def:ate}}]\\
&\le L\, W_1^{X,U}(S_{X,U}, Q_{X,U})
   && [\textrm{By \Cref{lem:LT}}] \\
&\le L\, \big[ W_1(S_X, Q_X) + \eta(S_U, Q_U) \big]
   && [\textrm{By \Cref{lem:glue}}] \\
\MSE_Q(S)
&\le L^2 \big[ W_1(S_X, Q_X) + \eta(S_U, Q_U) \big]^2 + \sigma^2_S
  \qedhere
\end{align*}
\end{proof}

\subsection*{A.7 Proof of \Cref{thm:alg}}\label{app:algorithms}

Recall that $\mathcal{Q}_h$ is the restriction of $\Theta_\rho$ to a grid. Define two optimal values, $V$, the optimal value over a directional shift family, and $V_h$, the optimal value over a grid with mesh size $h$:
\begin{equation*}
V = \min_{|S| = K} \sup_{\theta \in \Theta_\rho} W_1(Q_\theta, S_X),
\qquad
V_h = \min_{|S| = K} \max_{Q \in \cQ_h} W_1(Q, S_X).
\end{equation*}

\begin{lemma}[Grid approximation error]\label{lem:grid}
For every selection $S$,
\begin{equation*}
\max_{Q \in \cQ_h} W_1(Q, S_X)
\;\le\; \sup_{\theta \in \Theta_\rho} W_1(Q_\theta, S_X)
\;\le\; \max_{Q \in \cQ_h} W_1(Q, S_X) + c\,h ,
\end{equation*}
and consequently $V_h \le V \le V_h + c\,h$.
\end{lemma}

\begin{proof}
We show that approximating the parameter space by points on a grid adds error $ch$ to the upper bound minimized by \Cref{alg:formal}. 

For any $S$ and any $\theta, \theta'$:
\begin{align*}
W_1(Q_\theta, S_X) &\;\le\; W_1(Q_{\theta'}, S_X) + W_1(Q_\theta, Q_{\theta'}) &&[\textrm{Triangle inequality}]\\
&\;\le\; W_1(Q_{\theta'}, S_X) + c\,\lVert \theta - \theta' \rVert &&[\textrm{Lipschitz}]
\end{align*}
Every $\theta \in \Theta_\rho$ has a grid point $\theta'$ within $h$, and $\cQ_h \subseteq \{Q_\theta : \theta \in \Theta_\rho\}$, so
\begin{equation*}
\max_{Q \in \cQ_h} W_1(Q, S_X)
\;\le\; \sup_{\theta \in \Theta_\rho} W_1(Q_\theta, S_X)
\;\le\; \max_{Q \in \cQ_h} W_1(Q, S_X) + c\,h
\qquad \forall S .
\end{equation*}
Minimizing over $S$: $V_h \le V \le V_h + c\,h$.
\end{proof}

\begin{lemma}[The bounds bracket the optimum]\label{lem:bracket}
At each iteration $t$,
\begin{equation*}
\mathrm{LB}^{(t)} \;\le\; V_h \;\le\; \mathrm{UB}^{(t)} .
\end{equation*}
\end{lemma}

\begin{proof}
Recall that at each $t$ the algorithm stores a proposed site selection $\tS^{(t)}$, and a
collection $\tcQ^{(t)} \subseteq \cQ_h$ of adversarial shifts observed until $t$. From these it computes an upper bound and a lower bound on the worst-case site selection. 

First, the upper bound. The adversary evaluates the current site selection $\tS^{(t)}$ against every scenario on the grid and returns the worst observed loss. 
\begin{equation*}
\mathrm{UB}^{(t)} = \max_{Q \in \cQ_h} W_1\big(Q, \tS^{(t)}_X\big) .
\end{equation*}
$V_h$ is the same worst case minimized over \textit{all} selections, so it is smaller than the worst-case chosen by the adversary: 
\begin{equation*}
V_h = \min_{|S| = K} \max_{Q \in \cQ_h} W_1(Q, S_X)
\;\le\; \max_{Q \in \cQ_h} W_1\big(Q, \tS^{(t)}_X\big)
= \mathrm{UB}^{(t)} .
\end{equation*}

Next, we consider the lower bound. The planner solves a mixed-integer program whose value is the best
worst case over the scenarios found so far:
\begin{equation*}
\mathrm{LB}^{(t)} = \min_{|S| = K} \max_{Q \in \tcQ^{(t)}} W_1(Q, S_X) .
\end{equation*}
Now compare with $V_h$. For any fixed selection $S$, the scenarios in $\tcQ^{(t)}$ are a subset of the scenarios in $\cQ_h$, so:
\begin{equation*}
\max_{Q \in \tcQ^{(t)}} W_1(Q, S_X) \;\le\; \max_{Q \in \cQ_h} W_1(Q, S_X)
\qquad \text{for every } S 
\end{equation*}
And so, in particular:
\begin{equation*}
\min_{|S| = K} \max_{Q \in \tcQ^{(t)}} W_1(Q, S_X) \;\le\; \min_{|S| = K} \max_{Q \in \cQ_h} W_1(Q, S_X)
\end{equation*}
Where, plugging in definitions, we have:  
\begin{equation*}
\mathrm{LB}^{(t)} = \min_{|S| = K} \max_{Q \in \tcQ^{(t)}} W_1(Q, S_X) \;\le\;  \min_{|S| = K} \max_{Q \in \cQ_h} W_1(Q, S_X) = V_h .
\end{equation*}
\end{proof}

\begin{lemma}[Algorithm stopping rule]\label{lem:terminate}
\Cref{alg:formal} stops within $|\cQ_h|$ iterations, and when it stops,
$\min_{u < t} \mathrm{UB}^{(u)} - \mathrm{LB}^{(t)} \le \epsilon$.
\end{lemma}

\begin{proof}
Recall that at each iteration $t + 1$, the worst-case perturbation $\tilde{Q}^{(t+1)}$ is added to the collection, $\tcQ^{(t+1)} = \tcQ^{(t)} \cup \{\tQ^{(t+1)}\}$.

The planner then solves the master problem over the enlarged collection, and the value it attains is a new lower bound over this larger set:
\begin{equation*}
\mathrm{LB}^{(t+1)} = \max_{Q \in \tcQ^{(t+1)}} W_1\big(Q, \tS^{(t+1)}_X\big)
= \min_{|S| = K} \max_{Q \in \tcQ^{(t+1)}} W_1(Q, S_X) .
\end{equation*}

The upper bound at iteration $t$ is the distance from the current selection to the worst
scenario the adversary found for it, searching the whole grid:
\begin{equation*}
\mathrm{UB}^{(t)} = W_1\big(\tQ^{(t+1)}, \tS^{(t)}_X\big) = \max_{Q \in \cQ_h} W_1\big(Q, \tS^{(t)}_X\big) .
\end{equation*}
The stopping rule is that the best upper bound seen so far is within $\epsilon$ of the current
lower bound:
\begin{equation*}
\min_{u \le t} \mathrm{UB}^{(u)} - \mathrm{LB}^{(t+1)} \;\le\; \epsilon .
\end{equation*}
We show that this condition must be met within $|\cQ_h|$ iterations for any $\epsilon$. The basic idea is that the gap is exactly zero
whenever the adversary's best response to the planner is an adversarial shift that the planner has already observed. 

Suppose that $\tQ^{(t+1)} \in \tcQ^{(t)}$. Then, the planner solves the same master problem as before, so that:
\begin{equation*}
\mathrm{LB}^{(t+1)} = \mathrm{LB}^{(t)} 
\end{equation*}
Now compare the two bounds. Both are worst cases of the same selection $\tS^{(t)}$, but over different sets of scenarios: the upper bound over the whole grid, the lower bound over the
collection found so far:
\begin{equation*}
\mathrm{UB}^{(t)} = \max_{Q \in \cQ_h} W_1\big(Q, \tS^{(t)}_X\big),
\qquad
\mathrm{LB}^{(t)} = \max_{Q \in \tcQ^{(t)}} W_1\big(Q, \tS^{(t)}_X\big) .
\end{equation*}
The scenario that attains the first maximum is $\tQ^{(t+1)}$. If it already lies in
$\tcQ^{(t)}$, then it is also the lower bound, and hence: 
\begin{equation*}
\mathrm{UB}^{(t)} = \mathrm{LB}^{(t)}.
\end{equation*}
Which implies that $\min_{u \le t} \mathrm{UB}^{(u)} - \mathrm{LB}^{(t+1)} \le \mathrm{UB}^{(t)} - \mathrm{LB}^{(t)} = 0 \le \epsilon$, and hence the algorithm stops. 

This is guaranteed to occur, since the parameter space contains finitely many points. Otherwise $|\tcQ^{(t+1)}| = |\tcQ^{(t)}| + 1$. Since
$|\tcQ^{(0)}| = 1$ and $\tcQ^{(t)} \subseteq \cQ_h$ for all $t$, this happens at most
$|\cQ_h| - 1$ times: the algorithm stops within $|\cQ_h|$ iterations.
\end{proof}

\begin{proof}[Proof of \Cref{thm:alg}]
Termination is characterized by \Cref{lem:terminate}. Let $t$ be the stopping iteration and $u^{\star} = \arg\min_{u < t} \mathrm{UB}^{(u)}$,
so $\Sdro = \tS^{(u^{\star})}$. Then:
\begin{align*}
\sup_{\theta \in \Theta_\rho} W_1\big(Q_\theta, S^{*}_{\mathrm{DRO},X}\big)
&\le \max_{Q \in \cQ_h} W_1\big(Q, S^{*}_{\mathrm{DRO},X}\big) + c\,h
    && [\textrm{\Cref{lem:grid}}] \\
&= \mathrm{UB}^{(u^{\star})} + c\,h
   && [\textrm{definition of } \mathrm{UB}] \\
&\le \mathrm{LB}^{(t)} + \epsilon + c\,h
   && [\textrm{\Cref{lem:terminate}}] \\
&\le V_h + \epsilon + c\,h
   && [\textrm{\Cref{lem:bracket}}] \\
&\le V + \epsilon + c\,h
   && [\textrm{\Cref{lem:grid}}]
\end{align*}
\end{proof}

\end{document}